# Composition-dependent Thin-film Synthesis of Layered Ternary Iron Nitrides Fe*M*$N_2$ (*M* = W, Mo)

Baptiste Julien[1]*, Yuwei Yang[2], Liam A. V. Nagle-Cocco[3], Nicholas A. Strange[3], Nicholas M. Bedford[4,5], Andriy Zakutayev[1]*

1. *Materials Science Center, National Laboratory of the Rockies, Golden, CO 80401, United States.*
2. *InnoCORE AI-CRED Institute, Korea Advanced Institute of Science and Technology (KAIST), Daejeon 34141, Republic of Korea.*
3. *Stanford Synchrotron Radiation Lightsource, Stanford University, Menlo Park, CA 94025, United States.*
4. *Department of Chemistry, Colorado School of Mines, Golden CO 80401, United States.*
5. *Energy and Environment Science & Technology Directorate, Idaho National Laboratory, Idaho Falls, ID, 83415, United States.*

* *email:* Baptiste.Julien@nlr.gov, Andriy.Zakutayev@nlr.gov

## Abstract

Ternary transition-metal nitrides with layered crystal structure host anisotropic bonding and reduced dimensionality that may lead unconventional electronic and magnetic behavior. The synthesis of these layered nitrides phases in thin-film form is difficult because reactive sputtering often favors metastable rocksalt-derived structures. Here we report the composition-dependent thin-film synthesis and characterization of structure and properties of layered ternary iron nitrides Fe*M*$N_2$ (*M* = W, Mo) with triangular Fe sublattice, prepared by reactive sputtering of amorphous precursors and post-deposition ammonia ($NH_3$) annealing that staggers atomic mixing of the element precursors and long-range crystallization of extended structure. We find that $FeWN_2$ maintains good phase purity across a broad composition range indicative of cation substitution in the layered structure, while $FeMoN_2$ only shows good phase purity in Fe-poor composition, with an optimum corresponding to $Fe_{0.8}Mo_{1.2}N_2$ stoichiometry. Azimuthal grazing-incidence wide-angle X-ray scattering (GIWAXS) analysis shows that in both material systems, Fe-rich films exhibit strong out-of-plane fiber textures, and synchrotron X-ray absorption spectroscopy (XAS) provides insight into local Fe state and environment in both material systems. Room-temperature magnetization measurements of $FeWN_2$ reveal that while the stoichiometric film shows a simple paramagnetic behavior, the Fe-poor films are weakly ferromagnetic. Together, these results suggest that off-stoichiometry and local disorder within the triangular Fe sublattice may partially relieve magnetic frustration in the layered framework, and highlight the strong coupling between composition, microstructure, and magnetic properties in layered nitride materials.

## Introduction

Transition-metal (TM) nitrides exhibit a broad range of functional properties including superconductivity [1,2], magnetism [3,4], catalytic activity [5,6], and high electronic conductivity [7,8], making them attractive for applications spanning hard coatings, electronics, and energy technologies. While many TM nitride thin-films crystallize in simple rocksalt-derived structures, such as TiN, ScN, NbN or TaN [9–12], some multivalent ternary TM nitrides with $ABN_2$ formula adopt hexagonal layered structures [13,14]. One example of such structure, referred to as 'rockseline' in literature [15–17] is comprised of alternating sheets of edge-sharing $AN_6$ octahedra and $BN_6$ trigonal prisms, forming triangular lattices, such as reported in $MgMoN_2$, $MgWN_2$ or $ScTaN_2$ [16,18,19] . This "natural superlattice" framework is a promising candidate to host anisotropic bonding, reduced dimensionality and unconventional electronic and magnetic behavior.

In this category of layered nitrides, $FeWN_2$ and $FeMoN_2$ are especially intriguing due to the presence of Fe layers, forming triangular lattices with bond distances close to that in metals [20], which may host a frustrated magnetic behavior [21]. Previous bulk studies have reported the synthesis of $FeWN_2$ and $FeMoN_2$ in bulk form [22,23], while theoretical and experimental investigations of related compounds have suggested that their structural, electronic, and magnetic properties are highly sensitive to cation ordering, vacancies, and local coordination environment. However, the composition effects on phase stability, structural accommodation mechanisms and magnetic properties of these layered nitrides remain poorly understood, especially in thin-film form where there are no existing reports in literature. In addition, the relationship between long-range crystallographic ordering and the local electronic structure of Fe within the layered framework has not yet been systematically investigated.

Achieving these layered nitrides phases in thin-films remains highly challenging. This is because reactive sputtering of TM ternary nitrides often favor metastable rocksalt-derived structures with cation disorder [16,24,25]. To address this metastability challenge, specific annealing routes are used to overcome energetic barriers and induce phase nucleation, offering a promising pathway toward the formation of layered ternary nitride phases from metastable precursor structures. Furthermore, combinatorial thin-film synthesis enables efficient mapping of composition-dependent phase formation and structural evolution across broad compositional ranges, which is important as the stable layered structure does not necessarily exist at the nominal stoichiometric ratio [19,26]. At the same time, synchrotron-based characterization techniques provide powerful insight into both the long-range and local structure of these materials. Grazing-incidence wide-angle X-ray scattering (GIWAXS) enables detailed analysis of crystallographic texture and orientation distributions that are difficult to access using conventional diffraction methods alone [27,28], while X-ray absorption spectroscopy (XAS) provides complementary information regarding the local coordination environment and electronic structure of elements within the material long-range structure [29,30].

In this work, we report on the combinatorial synthesis of $Fe_xM_{1-x}N$ ($M$ = W, Mo) thin films, where $x$ = Fe/(Fe+$M$) is the atomic cation fraction, using reactive sputtering. Post-deposition annealing in $NH_3$ transforms the as-grown amorphous film precursors into the layered $FeMN_2$-type phase with a high degree of phase purity. Using synchrotron grazing-incident wide-angle X-ray scattering (GIWAXS) and X-ray absorption spectroscopy (XAS), we investigate the composition-dependent evolution of phase formation, crystallographic texture, layer ordering, and local Fe coordination. We find that while $FeWN_2$ maintains its layered structure across a broad range of metal compositions with good phase purity, $FeMoN_2$ exhibits a substantially narrower stability window and the excess Fe precipitates to form secondary phases. Azimuthal GIWAXS analysis reveals a pronounced texture redistribution with composition in both systems. Comparative analysis of ordering-sensitive diffraction intensity ratios together with XAS measurements further highlights fundamentally different structural accommodation mechanisms in the Fe$M$N$_2$ layered framework. Magnetic measurements at room-temperature suggest that off-stoichiometry within the layered $FeWN_2$ framework may influence magnetic frustration within the triangular Fe lattice, motivating future investigations of composition-dependent magnetic ordering in these layered nitrides. These results demonstrate that $FeWN_2$ and $FeMoN_2$, despite being isostructural, exhibit fundamentally different structural accommodation mechanisms of compositional variations, and highlights the strong coupling between composition, microstructure, and magnetic properties in layered nitride thin films.

## Materials and Methods

Thin films of Fe-W-N and Fe-Mo-N were synthesized using radio-frequency (RF) reactive magnetron co-sputtering in a confocal geometry. 2" targets of Fe, W and Mo (99.95% purity) were co-sputtered in an Ar+$N_2$ atmosphere with an Ar/$N_2$ ratio of 8:1 sccm. The RF power of all source targets was set to 40 W (power density of 1.97 W/cm$^2$). The base pressure was maintained below $3 \times 10^{-7}$ Torr and the deposition pressure was set to 8 mTorr. Composition-graded films were deposited on stationary 2" square Si/SiN$_x$ substrates (50.8 mm × 50.8 mm) with the SiN$_x$ layer (~100 nm thick) acting as a barrier layer to prevent reaction with Si during following anneals. Depositions were performed without external heating. After each deposition, the Fe-W-N and Fe-Mo-N libraries were cut into stripes along the gradient direction films, loaded in a tube furnace, ramped to 650 °C in $N_2$ at 10 °C/min and annealed for 2 hours in flowing $NH_3$. The films were then cooled naturally in $N_2$. A schematic illustration of the synthesis route is presented in Figure 1a.

Energy-dispersive X-ray fluorescence (ED-XRF) was employed to characterize the metal composition, using a Bruker M4 Tornado equipped with a Rh source operating at 50 kV and 200 μA. XRF spectra were sequentially acquired across the combinatorial films with a spot size of 25 μm and an exposure time of 120 s, and the metal ratio, defined as $x$ = Fe/(Fe+*M*) in at.%, where *M* = W or Mo, was quantified. Qualitative analysis of the nitrogen concentration before and after $NH_3$ annealing was performed by wavelength-dispersive X-ray fluorescence (WD-XRF) on a Rigaku ZSX Primus IV equipped with a Rh source operating at 30 kV and 100 μA . Semi-quantitative standardless analysis (SQX) was used to quantify the anionic ratio N/(O+N) before and after annealing

Laboratory X-ray diffraction (XRD) was used for quickly screening the as-grown films, using a Bruker D8 diffractometer equipped with an HI-STAR area detector. High-quality structural characterization was performed with synchrotron grazing-incidence wide-angle X-ray scattering (GIWAXS) at the Stanford Synchrotron Radiation Lightsource (SSRL beamline 11-3), using a 12.7 keV radiation source (λ = 0.97625 Å) and a large Rayonix MX225 CCD area detector. Diffraction patterns were collected with a 3° incident angle, a sample-to-detector distance of 150 mm, and a spot size of 50 μm by 150 μm. Two-dimensional reduction and azimuthal analysis of the GIWAXS data were performed using the NIKA package [31]. LeBail refinements were performed using GSAS II [32].

X-ray absorption (XAS) measurements were performed at the 10-ID-B beamline of the Advanced Photon Source, Argonne National Laboratory. Samples were mounted on sample holders positioned 45 degrees from the incident beam to collect data in a fluorescence geometry. XAS spectra were acquired at three different locations on the composition-graded films (center and edges), corresponding to Fe-rich ($x$ > 0.5), near-stoichiometric ($x \approx 0.5$), and *M*-rich ($x$ < 0.5) compositions. Measurements were taken from ~200 eV below to ~800 eV each edge. Data was processed using Athena, part of the Demeter software package [33]. The wavelet-transform was performed using the HAMA Fortran program [34]. Reference Fe K-edge XANES spectra used for oxidation-state comparison were obtained from the BL14B2 XAFS Standard Sample Database [35].

The magnetic response of the films at selected compositions was measured in a Quantum Design DynaCool Physical Property Measurement System (PPMS). Field-dependent magnetization loops were recorded from -20 kOe to 20 kOe at room-temperature. The substrate contribution was subtracted by measuring the bare substrate in the same conditions. More details about the processing of magnetic data employed in this work can be found in our previous work [36].

## Results and Discussion

### Thin-film synthesis of Fe(W,Mo)$N_2$

Combinatorial deposition of Fe-W-N and Fe-Mo-N in dilute nitrogen atmosphere (Ar:$N_2$ = 8:1) at room-temperature yielded films with composition ranges (defined as $x$ = Fe/Fe+$M$, where $M$ = W, Mo) of $x$ = 0.28 - 0.65 and $x$ = 0.32 - 0.69, respectively. The absence of diffraction peaks indicates that the as-grown films are amorphous or at least nanocrystalline, as suggested by the broad humps around 40° (Figure S1) probably from short-range ordering from metal-nitrogen bonds. The hump is more intense in Fe-W-N, likely due to the greater scattering amplitude of W compared to Mo.

The combinatorial films were then annealed at 650°C in flowing $NH_3$ for 2h. To evaluate the effect of $NH_3$ annealing on the nitrogen incorporation, WD-XRF measurements were performed on the as-grown and annealed Fe-W-N and Fe-Mo-N films, around $x$ = 0.5. WD-XRF spectra are shown in Figure 1(b,c). Nitrogen is clearly detected in the as-grown films whereas the oxygen signal remains relatively small. The anionic atomic ratio, defined as N/(O+N), was estimated by a standardless semi-quantitative analysis (SQX) and the results are presented in Table I. The ratio is similar in as-grown Fe-W-N (87.8 ± 2.6%) and Fe-Mo-M (85.4 ± 2.0%), suggesting that no significant oxidation occurred during or after the film deposition. After $NH_3$ annealing, both material systems exhibit a pronounced increase in the N-$K_\alpha$ line intensity, indicative of an effective nitrogen incorporation. The O-$K_\alpha$ intensity remains weak and shows little change after annealing, suggesting that the annealing process at 650 °C under flowing $NH_3$ prevent oxygen incorporation. Post-anneal anionic ratio are 95.2 ± 1.0% and 93.5 ± 0.9% in Fe-W-N and Fe-Mo-N film, respectively, confirming minor oxygen presence. Similar measurements were performed at different compositions $x$ and no significant changes on the N and O signal were observed (Figure S2), suggesting the local cation composition does not significantly affect the nitrogen incorporation or oxidation during $NH_3$ annealing.

**Table I**. Standardless semi-quantitative (SQX) estimation of the anionic ratio of Fe-W-N and Fe-Mo-N films measured by WD-XRF, before and after $NH_3$ annealing.

| **Film** | **N/(N+O) by SQX (at.%)** | |
|---|---|---|
| | **As-grown** | **$NH_3$ annealed** |
| Fe-W-N | 87.8 ± 2.6 | 95.2 ± 1.0 |
| Fe-Mo-N | 85.4 ± 2.0 | 93.5 ± 0.9 |

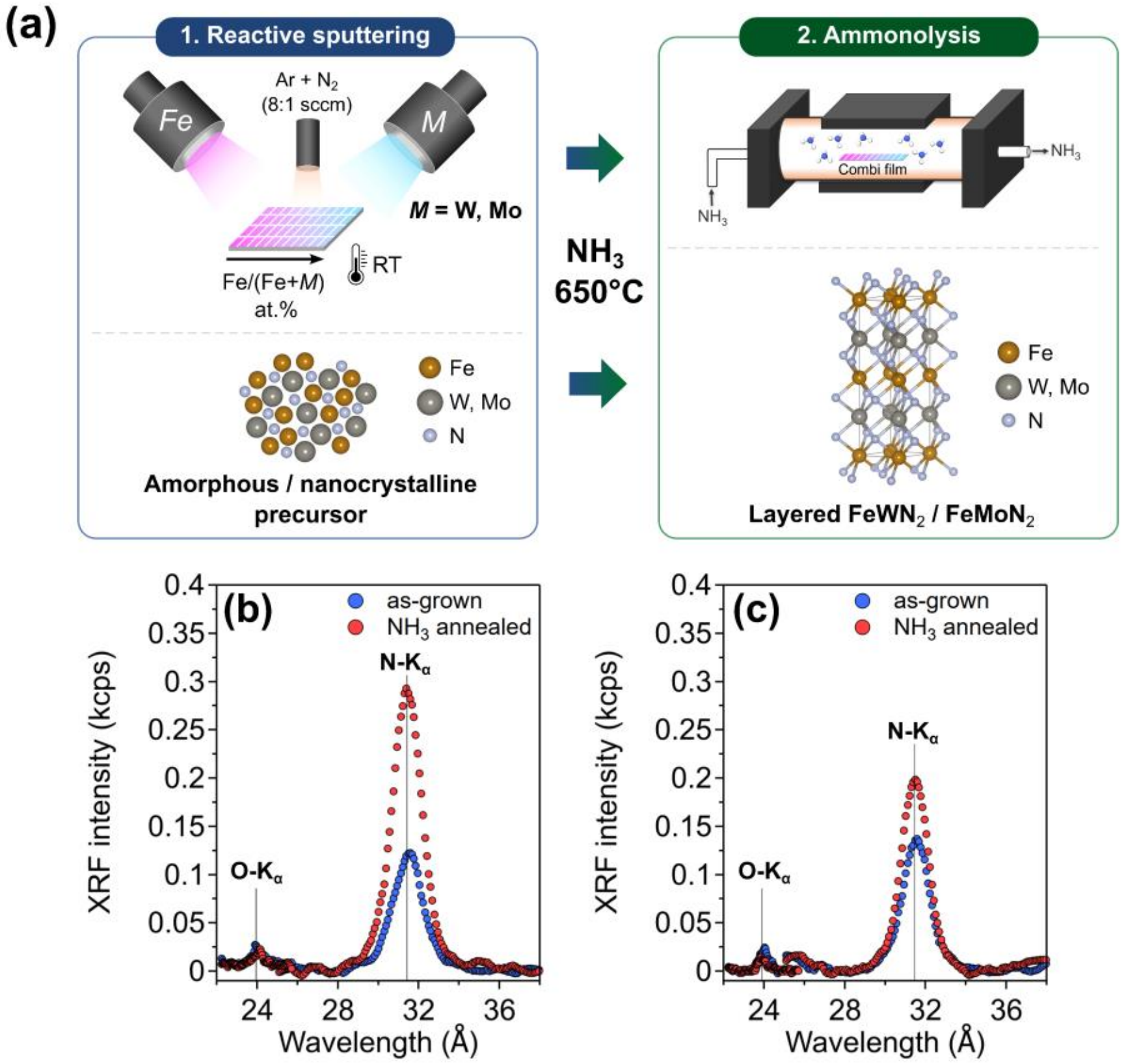

**Figure 1.** (a) Schematic illustration of the two-step synthesis process. Composition-spread amorphous (or nanocrystalline) Fe-M-N precursor films (M = W, Mo) are deposited by reactive sputtering and subsequently converted into layered $FeMN_2$ by $NH_3$ annealing. (b,c) WD-XRF spectra of Fe-W-N and Fe-Mo-N films around N-$K_\alpha$ and O-$K_\alpha$ emission lines, before and after $NH_3$ annealing,

## Crystal structure and phase purity

Figure 2(a,b) summarizes the results of GIWAXS measurements of the crystal structure on composition-graded Fe-W-N and Fe-Mo-N films annealed at 650°C in $NH_3$, respectively. Both W and Mo-based resulting films are polycrystalline and exhibit the $ABN_2$-type layered phase across the range of composition investigated. Layer ordering is well confirmed by the presence of the superlattice peak (002) at Q = 1.15 $Å^{-1}$ in both $FeWN_2$ and $FeMoN_2$. Interestingly for the former, the relative intensities of the different peaks change with composition and differ from the simulated diffraction patterns, suggesting composition-dependent crystallographic texture. The persistence of $FeWN_2$ structure for a broad range of composition indicates the layered structure can accommodate off-stoichiometry, likely through Fe/W substitutions. This has already been observed in isostructural nitrides [19,26,37,38]. For $FeMoN_2$, although the layered phase is observed at each composition, a fraction of $Fe_4N$ and α-Fe secondary phase increases as Fe content increases. We also note the presence of MoN at Mo-rich compositions.

The crystalline phase fraction was qualitatively estimated from XRD and defined as:

$$W_i(\%) = \frac{I_i^{exp}/I_i^{calc}}{\sum_j I_j^{exp}/I_j^{calc}} \tag{1}$$

where $W_i$ is the fraction of the phase $i$, and $I_i^{exp}$ and $I_i^{calc}$ are the experimental and calculated intensities of a characteristic peak of phase $i$. Details about calculation of the phase fraction can be found in Table S1 and Table S2. The composition-dependent phase fractions for Fe-W-N and Fe-Mo-N films are presented in Figure 2(c,d). The Fe-W-N system shows a large phase-pure window for $FeWN_2$ centered around stoichiometry, with a small fraction of $W_3N_4$ and $Fe_3N$ emerging at W-rich and Fe-rich compositions, respectively.

In contrast, in the Fe-Mo-N system, only a narrow composition region around $x$ = 0.4 shows the layered structure with phase purity (within diffraction limit), corresponding to $Fe_{0.8}Mo_{1.2}N_2$, assuming nitrogen stoichiometry. Above this Fe content, α-Fe and $Fe_4N$ fraction significantly increases to a point where $FeMoN_2$ fraction is only 53% at $x$ = 0.69. Thus, the layered phase observed is likely to be a Fe-defective $FeMoN_2$ structure. Below $x$ = 0.4, the fraction of MoN phase rapidly increases. Noteworthily, the phase-pure composition observed in this work corresponds to the one successfully synthesized and characterized in bulk form as reported in literature [23]. These observations suggest that the layered phase is intrinsically stabilized under Fe-deficient/Mo-rich conditions rather than at ideal $FeMoN_2$ stoichiometry, while Fe-rich compositions excess the solubility window and form secondary α-Fe and $Fe_4N$ phases, likely due to the slight thermodynamic metastability of $FeMoN_2$ [36].

It must be noted that the two other competing ternary phases with the same 1:1 metal ratio $Fe_3W_3N$ and $Fe_3Mo_3N$, crystallizing in the η-carbide type structure (*Fd-3m*), were not formed here. In a recently published work, we were able to successfully synthesize these N-poor phases in thin-films using rapid thermal annealing under flowing $N_2$ [36]. The greater chemical potential provided by the $NH_3$ atmosphere, compared to $N_2$, likely favors the formation of the N-rich phases $FeMN_2$ rather than N-poor $Fe_3M_3N$ phases, at least when the initial precursor film is amorphous. This provides interesting insights into thin-film synthesis strategies to navigate into complex nitride landscapes.

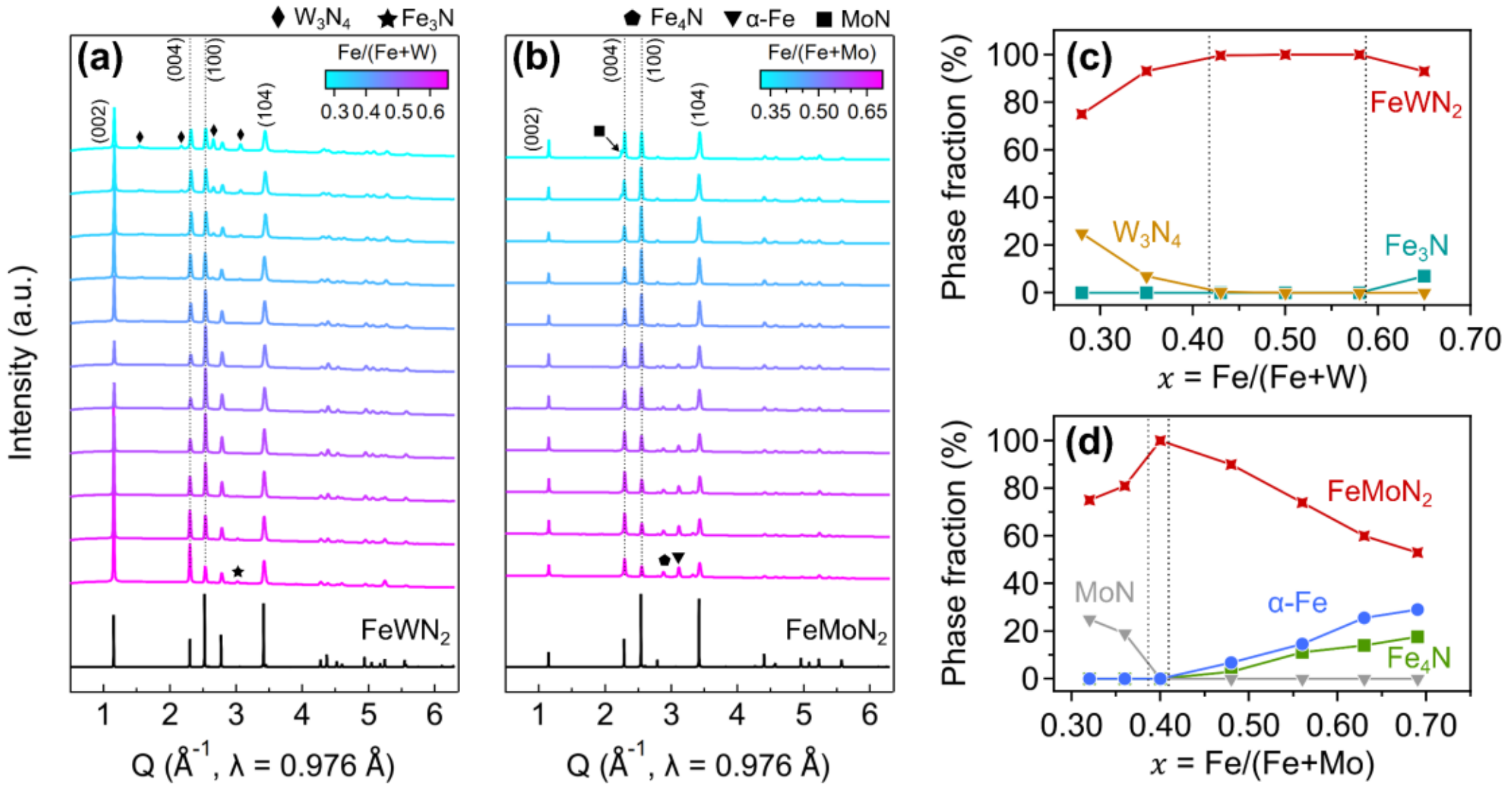

**Figure 2**. Composition-dependent GIWAXS patterns of composition-graded (a) Fe-W-N and (b) Fe-Mo-N films collected at λ = 0.976 Å. Traces are vertically offset for clarity. Simulated patterns of $FeWN_2$ and $FeMoN_2$ are shown for reference. Peaks corresponding to secondary phases are marked with symbols. Composition-dependent estimated phase fraction for (c) Fe-W-N and (d) Fe-Mo-N films. Gray dash lines highlight the composition windows in which phase-pure $FeWN_2$ and $FeMoN_2$ is observed.

## Lattice parameters and layer ordering

The lattice parameters of $FeWN_2$ and $FeMoN_2$ phases were estimated from (004) and (104) peaks and their evolution with composition is displayed in Figure 3(a,b). The values of $a$ and $c$ for $FeMoN_2$ at $x$ = 0.4 (corresponding to $Fe_{0.8}Mo_{1.2}N_2$) match perfectly the bulk results from literature [23]. The relatively small variation of $a$ and $c$ with composition in $FeMoN_2$ is likely due to phase segregation. In contrast, a sharp transition in the $c$ parameter in $FeWN_2$ is observed around stoichiometry ($x$ ~ 0.5) where the c-axis rapidly expands from 10.88 Å to 10.96 Å between $x$ = 0.46 and $x$ = 0.55. This cannot be explained by a simple gradual Fe/W substitution behavior, but instead could be due to an abrupt change in compressive strain-induced texture around stoichiometry. The value of $c$ at $x$ = 0.5 is close to the reported bulk value [22].

The degree of layer ordering was investigated by comparing the (002) superlattice reflection to the (004) fundamental reflection. Because both the (002) and (004) reflections probe the same crystallographic direction, their relative integrated intensity is expected to be comparatively insensitive to first-order texture evolution, but sensitive to layer ordering and stacking coherence along the c-axis. To minimize geometric bias associated with the GIWAXS missing wedge along the out-of-plane axis, diffraction intensities were integrated within an azimuthal sector centered at 35° with an azimuthal width of 10°, as illustrated in Figure S3. Under these conditions, variations in $I_{002}/I_{004}$ ratio can be interpreted primarily as changes in layer ordering and stacking coherence rather than reciprocal-space accessibility effects.

Figure 3(c,d) shows the composition dependence of $I_{002}/I_{004}$ for $FeWN_2$ and $FeMoN_2$ films, respectively. In $FeWN_2$, the ratio progressively increases with Fe content, with a maximum at $x$ = 0.65 and suggests an enhanced stacking coherence along the $c$-axis at Fe-rich composition. This coincides with the strong out-of-plane texture and expansion of the $c$ lattice parameter, suggesting a composition-driven reorganization of the layered $FeWN_2$ structure. As a deviation from a 1:1 cation ratio should negatively impact the $I_{002}/I_{004}$ ratio, the strong $c$-axis texture could favor layer-by-layer nucleation, potentially reduce antisite disorder or stacking faults, and result in a more coherent layered structure overall, despite the deviation from stoichiometry. In contrast, $FeMoN_2$ exhibits a decreasing ratio with increasing Fe content, indicating progressive disruption of the layered stacking periodicity. This behavior is consistent with the narrow stability window of the layered phase near $x$ = 0.4 beyond which excess Fe precipitates and form secondary phases.

For comparison, the theoretical $I_{002}/I_{004}$ ratios were calculated from ideal structures generated in CrystalDiffract®. The experimental and calculated ratios are summarized in Table II for selected compositions. For stoichiometric $FeWN_2$, the small deviation from the calculated value indicates a certain degree of disorder from antisite defects, stacking faults or nitrogen vacancies, as it is often observed in layered nitrides and oxides [13,39]. Surprisingly, experimental $I_{002}/I_{004}$ of $Fe_{0.80}Mo_{1.20}N_2$ ($x$ = 0.4) matches perfectly the calculated one, indicating identical stacking coherence and suggesting again that the layered phase $FeMoN_2$ is intrinsically stabilized under Mo-rich conditions (at 0.8:1.2 metal ratio) rather than at nominal stoichiometry (1:1).

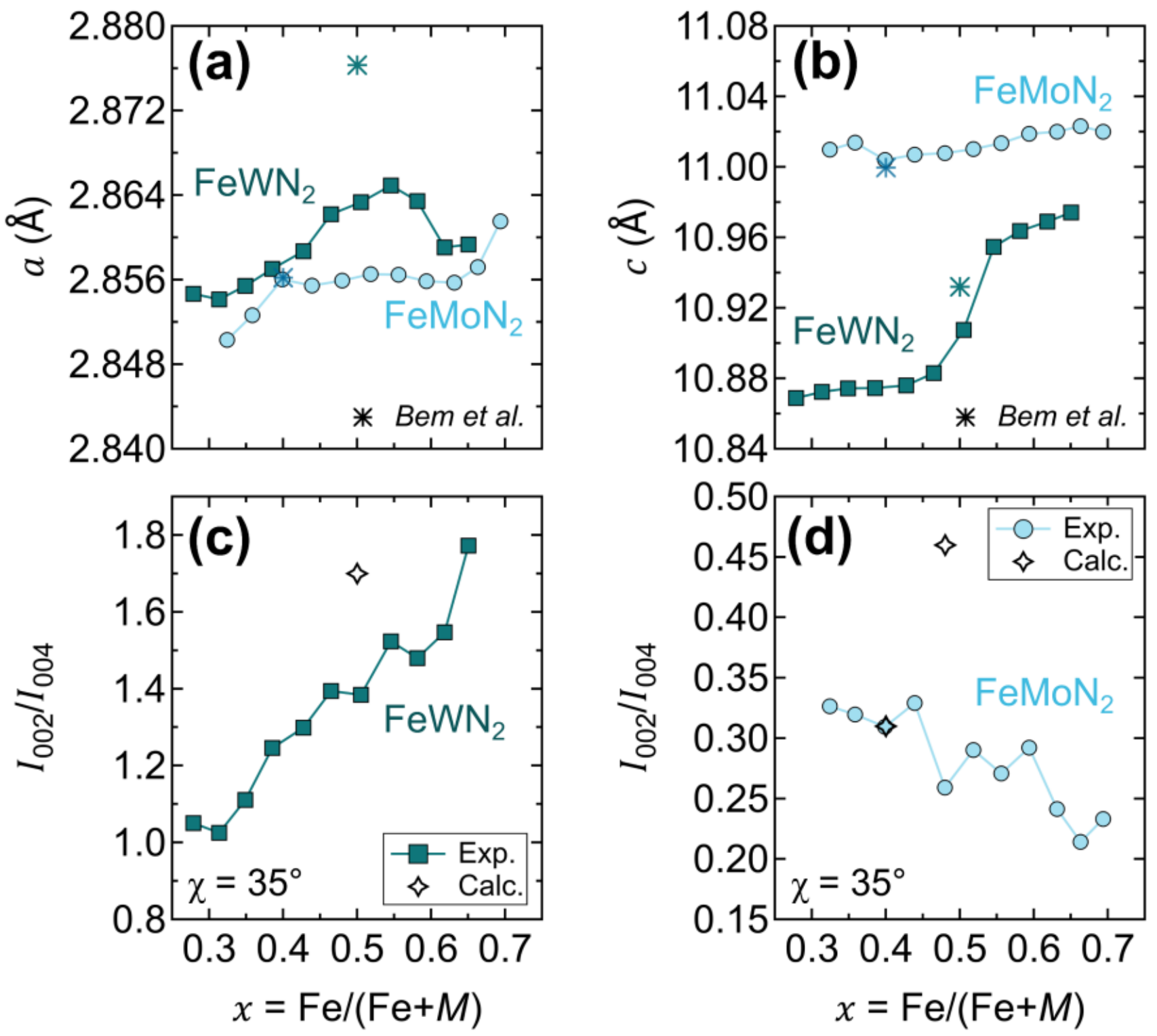

**Figure 3**. (a,b) Evolution of the lattice parameters $a$ and $c$ with the Fe atomic fraction $x$ = Fe/(Fe+*M*). The asterisk symbol denotes reported bulk values from literature for stoichiometric $FeWN_2$ and $Fe_{0.8}Mo_{1.2}N_2$ [22,23]. (c,d) Composition dependence of the integrated intensity ratio $I_{002}/I_{004}$ extracted from sector integrations of GIWAXS images at $\chi$ = 35° (with an azimuthal width of 10°). Dark symbols correspond to the calculated ratios at the compositions discussed in the text.

**Table II**. Experimental and calculated diffraction intensity ratios $I_{002}/I_{004}$ of $FeWN_2$ and $FeMoN_2$ structures calculated at different compositions.

| **Composition** | **x = Fe/(Fe+*M*)** | **Experimental $I_{002}/I_{004}$** | **Calculated $I_{002}/I_{004}$** |
|---|---|---|---|
| $FeWN_2$ | 0.50 | 1.38 | 1.70 |
| $Fe_{0.96}Mo_{1.04}N_2$ | 0.48 | 0.28 | 0.46 |
| $Fe_{0.80}Mo_{1.20}N_2$ | 0.40 | 0.31 | 0.31 |

* Structures are generated from cif files with ICSD entries for $FeWN_2$ (ICSD 81488) and for $Fe_{0.80}Mo_{1.20}N_2$ (ICSD 80892). Off-stoichiometry structures are generated by changing Fe/Mo site occupancy, assuming the nitrogen occupancy of 1. Diffraction intensities are calculated in CrystalDiffract® with λ = 0.97625 Å.

**Structure refinement**

For a more complete crystallographic analysis, GIWAXS patterns of $FeWN_2$ and $FeMoN_2$ films near $x$ ~ 0.5 were further refined using a LeBail method. Figure 4 presents the results of refinements and the corresponding GIWAXS images converted in reciprocal space with Ewald sphere correction accounting for the missing wedge in the out-of-plane direction high angles. The $Fe\mathit{M}N_2$ structure ($M$ = W, Mo), consisting of alternating layers of $FeN_6$ octahedra and $\mathit{M}N_6$ trigonal prisms, is shown Figure 4e. Figure 4f presents a plane view of the Fe layer, illustrating the triangular lattice.

For $FeWN_2$, the diffraction pattern is well fitted using a single-phase model of $P6_3/mmc$ $FeWN_2$ and the resulting fit is in excellent agreement between experimental data and calculated profile, with minimal residuals ($R_{wp}$ = 3.80%), indicating high phase purity and good crystallinity. The lattice parameters extracted by the LeBail method are $a$ = 2.86316(18) Å and $c$ = 10.9129(6) Å. The absence of secondary peaks confirms that the synthesis conditions stabilize the $FeWN_2$ phase at this composition without formation of competing phases.

In the case of $FeMoN_2$, a single-phase model is no longer sufficient to fit the whole pattern, especially in the Q = 2.7 – 3.2 $Å^{-1}$ region where extra peaks from secondary phases are present due to phase segregation. Instead, a multi-phase fit adding $Fe_4N$ (*Pm-3m*) and α-Fe (*Im-3m*) is employed to account for all observed reflections. The peaks at 2.88 $Å^{-1}$ and 3.11 $Å^{-1}$ correspond to (111) and (110) reflections of $Fe_4N$ and α-Fe, respectively. The resulting fit yields lattice parameters of $FeMoN_2$ phase $a$ = 2.85182(19) Å and $c$ = 10.9956(4) Å, with $R_{wp}$ = 4.40%. The extracted lattice parameters of secondary α-Fe and $Fe_4N$ are $a$ = 2.85221(17) Å and $a$ = 3.7409(14) Å, respectively, which is relatively close to bulk reported values. The non-uniform intensity distribution along the Debye rings on the associated 2D images of $FeWN_2$ and $FeMoN_2$, shown in Figure 4(c,d), indicate a degree of preferential orientation, which will be further analyzed in the next section.

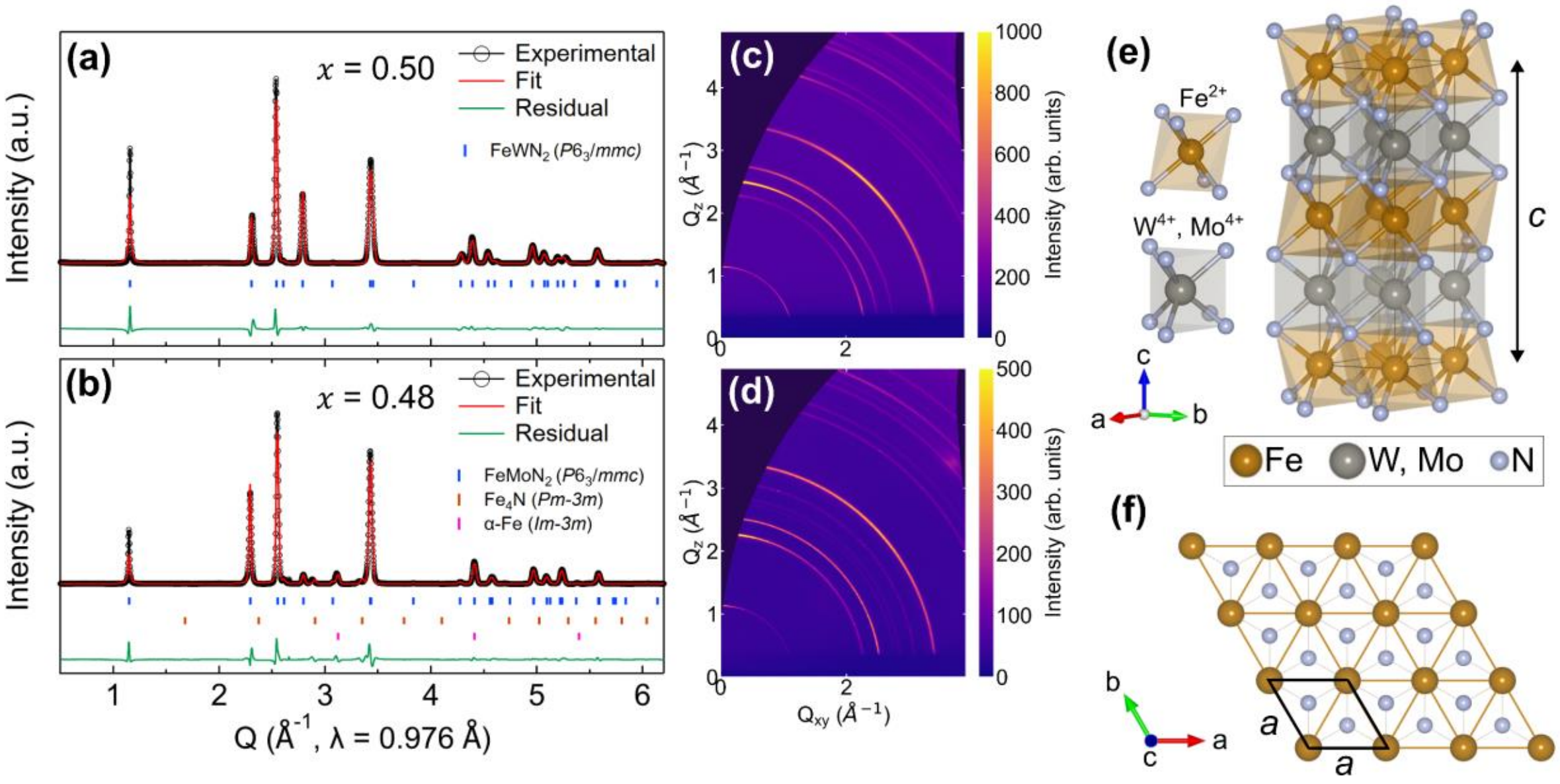

**Figure 4.** (a,b) LeBail refinements of GIWAXS patterns (full integration) of near-stoichiometric films $FeWN_2$ (x = 0.50) and $FeMoN_2$ (x = 0.48), respectively, into the hexagonal layered phase structural model (space group $P6_3/mmc$). Experimental data are shown as black circles, fit as red lines, and residuals as green lines. Tick marks indicate the expected Bragg reflection positions for the considered phases, including secondary phases $Fe_4N$ and α-Fe in $FeMoN_2$. (c,d) Corresponding GIWAXS images of converted in reciprocal space. (e) Crystal structure of layered $Fe\textit{M}N_2$. (f) Top view of the Fe triangular lattice.

### Crystallographic orientation

The crystallographic texture in composition-dependent $FeWN_2$ and $FeMoN_2$ films was investigated through azimuthal intensity analysis of the GIWAXS images. Representative reciprocal-space diffraction images for Fe-rich and near-stoichiometric compositions are shown in Figure 5(b,d), while the corresponding azimuthal intensity distributions extracted from the characteristic (002), (100), and (104) reflections are summarized in Figure 5(a,c). The gray shaded regions around $\chi$ = 90° correspond to the intensity affected by incomplete reciprocal-space accessibility and therefore cannot be quantitatively interpreted.

For both $FeWN_2$ and $FeMoN_2$, Fe-rich films exhibit a pronounced preferential orientation characterized by a strong maximum of the (002) reflection centered around $\chi$ = 90°, indicating that the basal plane is preferentially oriented parallel to the surface. In this regime, the (104) reflection displays a characteristic double-peak distribution with maxima located symmetrically around $\chi$ ≈ 45° and 135°, whereas (100) progressively increase approaching the in-plane direction. These observations suggest a fiber texture distribution with the $c$-axis being the fiber axis, and where the angular relationship between the (104) and ($00l$) planes is preserved across the crystallite population. Additional pole figures were performed on Fe-rich films to further visualize the fiber texture (Figure S4).

As the metal composition evolves toward $x$ = 0.5 in $FeWN_2$, the (100) azimuthal profile rapidly shifts to a broad peak centered around $\chi$ = 90°, accompanied by a drop in intensity of the (002) reflection. This indicates a redistribution of crystallographic orientations away from a single out-of-plane fiber texture toward a more predominant in-plane texture. We note that the (002) reflection maintains a small maximum around $\chi$ = 90° which is indicative of a broader mixed-orientation state. A careful analysis of the (104) azimuthal profile of $FeWN_2$ reveals a subtle shift of the maximum in $\chi$ of ~ 5° between $x$ = 0.65 and $x$ = 0.50 (Figure S5). This corresponds to the interplanar angle difference between (104)/(002) and (104)/(100) in $FeWN_2$ structure. Moreover, for $x$ = 0.50, the peak shows a shoulder aligning with the peak for $x$ = 0.35, indicative of a fraction of crystallite remaining in a $c$-axis fiber texture orientation.

In $FeMoN_2$, a similar texture is observed in Fe-rich films, although changes with composition appear more gradual, and the residual out-of-plane component remains more pronounced over the explored composition range. However, we expect that the phase mixture present at Fe-rich compositions due to phase segregation likely reduce the effect of the composition on the crystallographic texture of $FeMoN_2$ phase. Interestingly, around the $x$ = 0.4 composition, the film evolves to a more "powder-type" polycrystalline regime, indicated by isotropic GIWAXS rings (Figure S6).

In both systems, the evolution of the azimuthal distributions demonstrates that compositional tuning affects not only phase formation but also the orientation selection during crystallization of the layered phase. Overall, the GIWAXS azimuthal analysis reveals a continuous composition-driven texture reorientation in both layered nitride systems, with Fe-rich films favoring a well-defined out-of-plane fiber texture and compositions closer to stoichiometry exhibiting increasingly distributed and mixed crystallographic orientations.

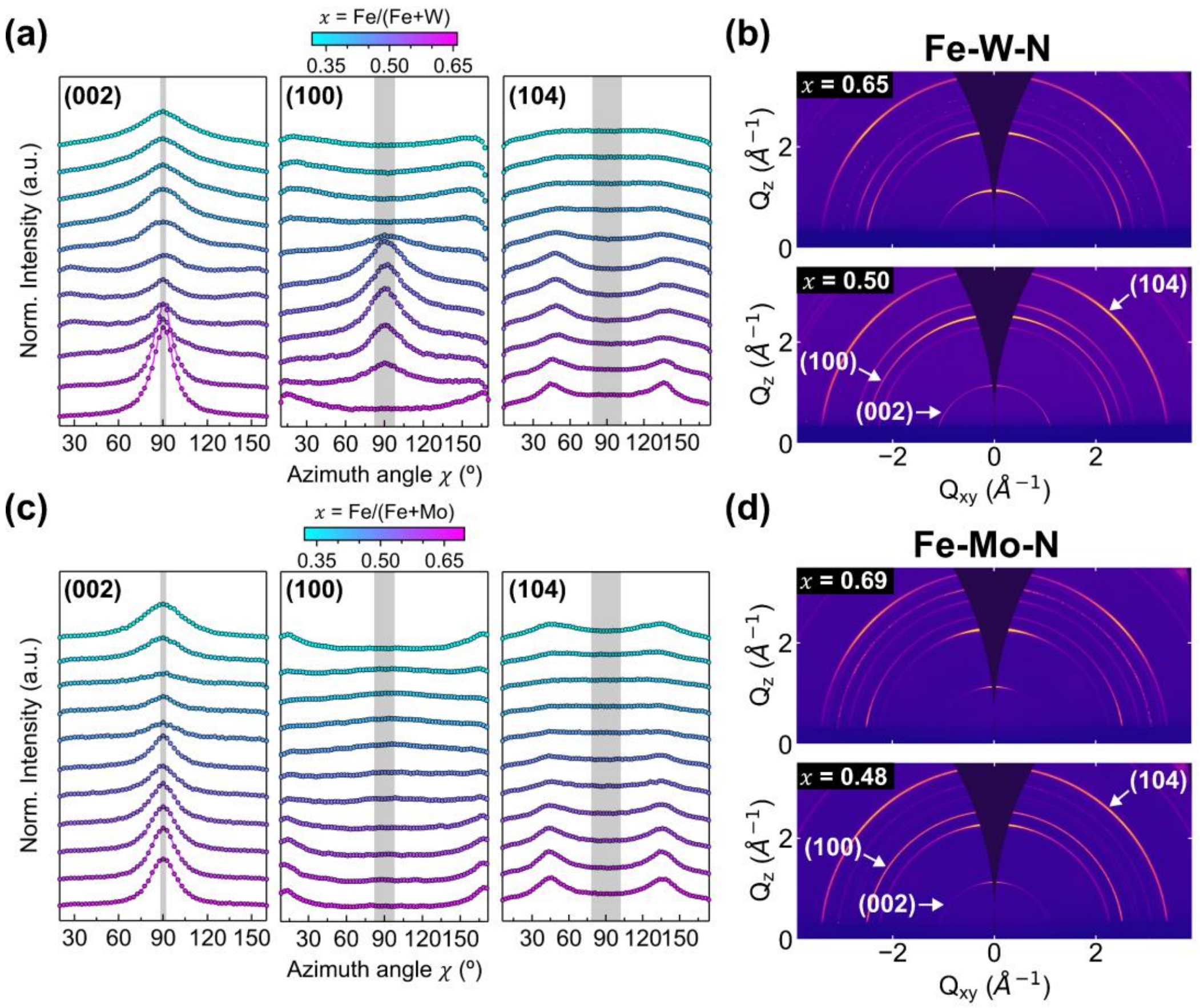

**Figure 5.** (a,c) Composition-dependent azimuthal intensity profiles extracted from GIWAXS images for the characteristic (002), (100), and (104) reflections of $FeWN_2$ and $FeMoN_2$ films, respectively. Curves are vertically offset for clarity and colored according to the metal ratio Fe/(Fe+*M*). The gray shaded regions around $\chi$ = 90° correspond to the intensity affected by incomplete reciprocal-space accessibility due to GIWAXS geometry. (b,d) Experimental GIWAXS reciprocal space maps for representative Fe-rich and near-stoichiometric compositions

in $FeWN_2$ and $FeMoN_2$ systems, respectively. The (002), (100) and (104) Debye-Scherrer rings are indicated for clarity.

## X-ray absorption spectroscopy

Fe K-edge XAS measurements were performed to investigate the evolution of the local Fe coordination environment as a function of composition in $FeWN_2$ and $FeMoN_2$ films. Figure 6(a,b) shows the Fe K-edge X-ray absorption near-edge structure (XANES) spectra of $FeWN_2$ and $FeMoN_2$, respectively. The Fe K-edge XANES spectra of both $FeWN_2$ and $FeMoN_2$ show an absorption edge with $E_0$ at ~7110 eV, slightly shifted to higher energy by 1-2 eV relative to the Fe foil. The edge remains below that of $Fe_2O_3$ and coincide well with FeO, broadly consistent with an oxidation state around 2+ (rather than 3+), as expected with a probable +4 oxidation state for W and Mo [40]. To further support this analysis, XANES spectra were compared with additional Fe-based standards including nitrides (Figure S7). A small systematic shift toward lower energy with increasing Fe content is observed, suggesting subtle modifications of the local Fe electronic environment rather than a significant change in the oxidation state. This shift is enhanced in $FeMoN_2$ system and is presumably due to the greater fraction of α-Fe and $Fe_4N$ phases at these compositions, lowering the effective electron deficiency around Fe.

A weak pre-edge feature is observed in $FeWN_2$ films, generally attributed to quadrupole 1s → 3d transitions. It is known that the pre-edge intensity is greatly enhanced by 3d-4p hybridization in non-centrosymmetric coordination which allows the electric-dipole 1s → 4p transition, and suggests that Fe exists in a distorted octahedral environment [41]. Moreover, its intensity decreases and shifts to lower energy as the Fe fraction $x$ increases, indicating a change in the local distortion around Fe atoms, that is likely caused by W substitution and the sudden change in the $c$ parameter (Figure 4b). In addition, the Fe-rich film ($x$ = 0.65) exhibits a less intense white-line, which is likely a consequence of a less electronegative environment due to W deficiency. In $FeMoN_2$ (Figure 6b), a notable lack of pre-edge feature is likely caused by limited 3d-4p hybridization as compared to $FeWN_2$ and more ordered Fe local structure, as evidence by GIWAXS results. The Fe-poor film ($x$ = 0.32) exhibits the sharpest white-line as well as the most pronounced post-edge oscillations, indicative of enhanced local structural coherence, consistent with GIWAXS results. The loss of intensity and shift of the white-line at higher Fe content is likely due to phase mixture and the lack of Mo polarization of Fe atoms.

Figures 6(c,d) present the real-space Fourier-transformed (FT) EXAFS data (non-phase-corrected) of $FeWN_2$ and $FeMoN_2$ films at the different compositions. Note that the FT-EXAFS spectra are not corrected for phase shift and are ~ 0.3 Å shorter than the real nearest neighbor distances. FT-EXAFS spectra of $FeWN_2$ films exhibit similar features across all stoichiometries, with variations in intensity and position, caused by the influence of metal composition on the local environment. The first peak at 1.45 Å (without phase correction) is assigned to Fe-N first shell, with a maximum intensity for the film close to stoichiometry ($x$ = 0.46). This is likely due to N vacancy formation to accommodate the excess Fe and associated domains at this stoichiometry. The second feature at ~1.9 Å are likely Fe-Fe/W overlapping distances. The relative intensity of these contributions and distance varies with Fe content, indicating composition-dependent changes in the local Fe coordination. The 2-3 Å region of the EXAFS is best described as overlapping Fe-W/Fe-Fe metal shell contributions due to composition-dependent variations in Fe/W site occupancy, with the high ($x$ = 0.65) and low ($x$ = 0.28) Fe stoichiometry materials exhibiting maximum that could correlate to likely Fe-Fe and Fe-W distance at ~2.3 and ~2.65 Å respectively. To support the analysis wavelet-transform EXAFS contour plots are provided in the Supplementary Material (Figure S8).

The FT-EXAFS of $FeMoN_2$ films in Figure 5d films reveal substantial changes across composition, likely dominated by the Fe-containing secondary phases Fe-rich compositions. Most notably is the broad distribution of Fe-N distances as compared to $FeWN_2$ films, strongly indicating and inability to accommodate off-stoichiometry phases. The Fe-poor film, which does not contain Fe-containing secondary phases, shows two main features at ~1.7 Å and ~2.5 Å which we can attribute to Fe-N and Fe-Mo, respectively.  These nearest neighbor distances agree with the theoretical structure and reported results [23]. It is also possible that Fe-Mo overlap with intralayer Fe-Fe contributions due to their similar distance.

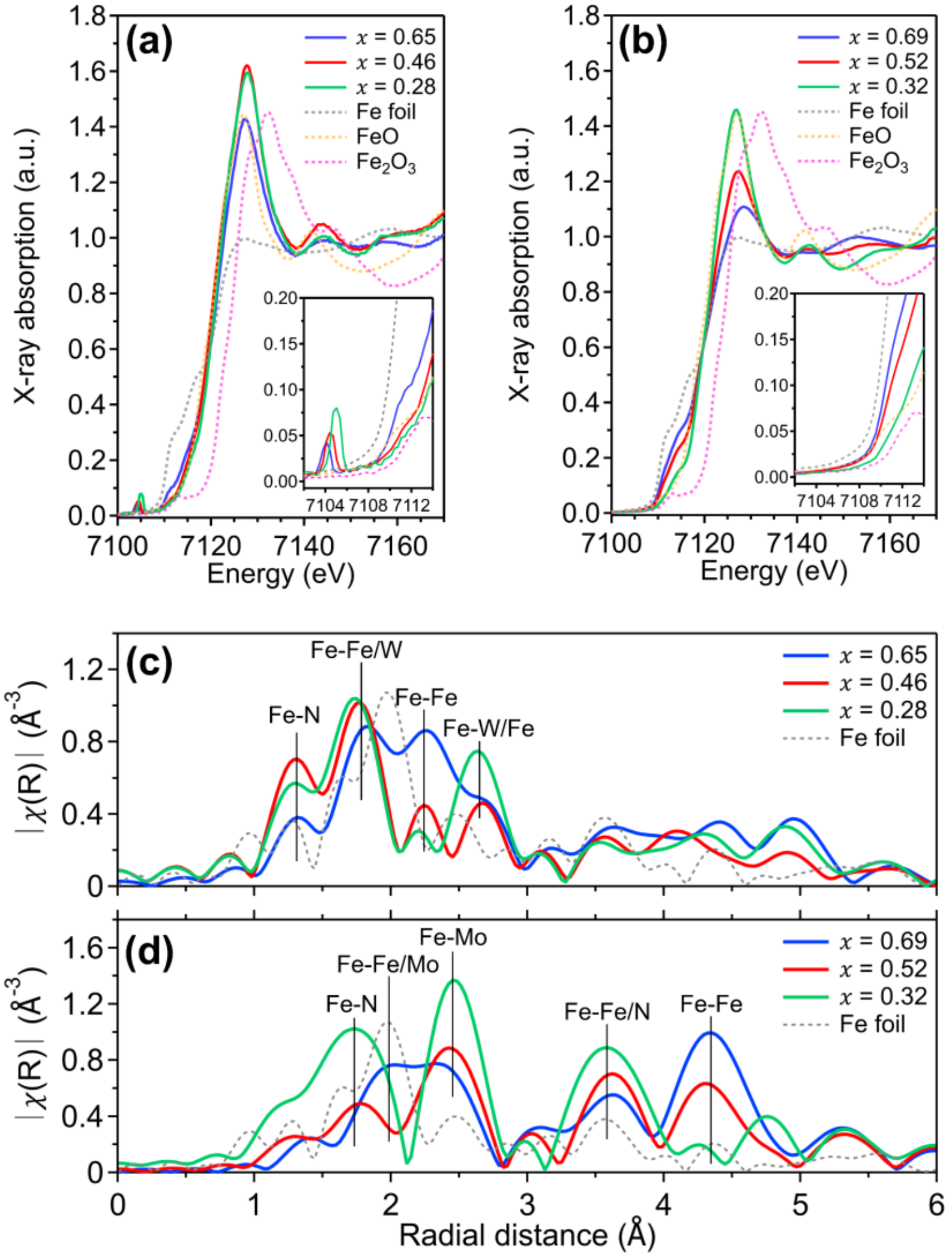

**Figure 6**. (a,b) Normalized Fe K-edge XANES spectra for $FeWN_2$ and $FeMoN_2$ films, respectively, at different Fe fraction $x$ corresponding to Fe-rich, near-stoichiometric and Fe-poor compositions. The insets highlight the pre-edge region. For oxidation state reference, XANES spectra of a Fe foil as well as FeO and $Fe_2O_3$ standards are added. (c,d) Magnitude of FT-EXAFS spectra (non-phase-corrected) for the corresponding $FeWN_2$ and $FeMoN_2$ films.

Additional wavelet-transform EXAFS contour plots are available in the Supplementary Materials (Figure S8).

**Magnetic properties**

To preliminarily explore the magnetic behavior of these ternary iron-based nitrides, room-temperature magnetization measurements were performed on stoichiometric $FeWN_2$ ($x$ = 0.5), $Fe_{0.76}W_{1.24}N_2$ ($x$ = 0.38) and $Fe_{0.8}Mo_{1.2}N_2$ ($x$ = 0.4) films, all featuring phase purity with respect to GIWAXS. As shown in Figure 7a, while stoichiometric $FeWN_2$ exhibits a paramagnetic-like linear response, the Fe-poor $Fe_{0.76}W_{1.24}N_2$ film shows an interesting non-linear magnetic response with a transition around 1 kOe. The near-zero coercivity and the magnetization rapidly reaching saturation are indicative of a soft ferromagnetic ordering, comparable in magnitude to previously reported values. This strong composition dependence suggests that the magnetism is closely linked to the structural accommodation mechanisms in the layered framework rather than arising solely from ferromagnetic impurities as no secondary phases, crystalline at least, were detected in GIWAXS at these compositions.

On the other hand, an hysteresis loop is observed at room-temperature in $Fe_{0.8}Mo_{1.2}N_2$ (Figure 7b), with a coercivity around 400 Oe, which is likely to be a signature of ferromagnetic impurities. Although no crystalline α-Fe or Fe-N phases could be identified in GIWAXS at this composition, because of the narrow phase-pure window, it is possible than the larger sample (~5 mm x 5 mm) measured by magnetometry contains few percent of ferromagnetic α-Fe and/or $Fe_4N$. However, the electrical resistivity of $FeMoN_2$ exhibits a pronounced resistivity maximum near nominal stoichiometry, coinciding with enhanced structural disorder and phase competition (Figure S9), rather than changing with Fe composition which would be expected if α-Fe and/or $Fe_4N$ presence were significant. For comparison, the electrical resistivity of $FeWN_2$ is relatively low (~1 mΩ.cm) and composition-independent.

The absence of magnetic response at room temperature in stoichiometric $FeWN_2$ is possibly caused by antiferromagnetic interactions within the frustrated triangular Fe lattices in the ordered structure. In fact, it was shown that $FeWN_2$ orders antiferromagnetically around 45 K in bulk [42], while the possibility of spin-glass behavior at low temperature was also reported [43]. Small disruption in the structural ordering in Fe-poor $Fe_{0.76}W_{1.24}N_2$, with W substitution and/or Fe vacancies in the octahedral layer can partially relieves magnetic frustration, stabilizing a weak ferromagnetic or canted magnetic state in Fe-deficient films, as already observed in bulk [43]. Similar antiferromagnetic ordering was reported in bulk $Fe_{0.8}Mo_{1.2}N_2$, and it is expected that both material order antiferromagnetically at low temperature [23]. Additional measurements and further investigations would be needed to confirm these hypotheses. These preliminary results suggest that magnetic ordering in $FeWN_2$ thin films may be highly sensitive to composition and local structural disorder, motivating future temperature-dependent and element-specific magnetic investigations. For $FeMoN_2$, it is hard to conclude as the main challenge remains in controlling phase-purity in thin-film form, to isolate the magnetic contribution from the layered structure.

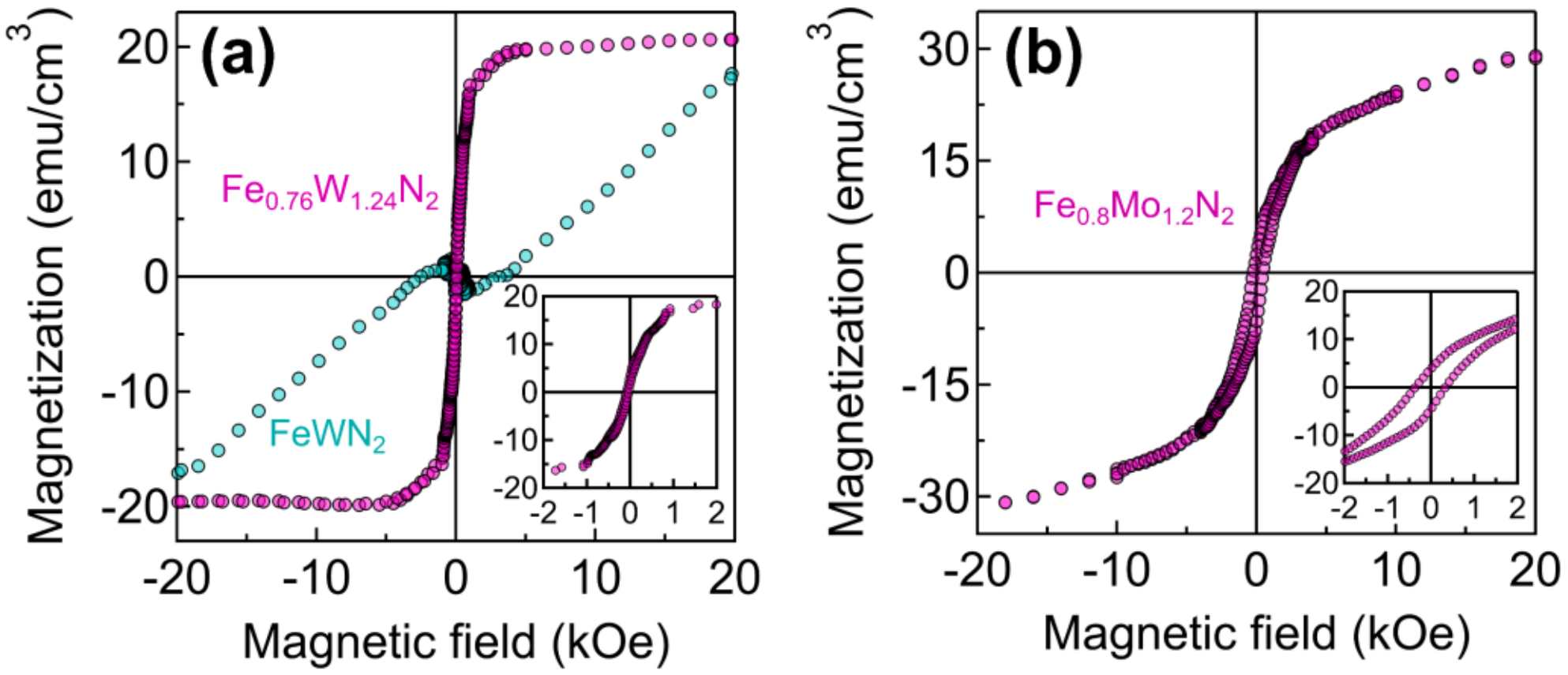

**Figure 7.** Room-temperature magnetization loops measured for phase-pure films with different Fe fractions: (a) stoichiometric $FeWN_2$ and Fe-poor $Fe_{0.76}W_{1.24}N_2$, and (b) Fe-poor $Fe_{0.8}Mo_{1.2}N_2$. The contribution from the Si/$SiN_x$ substrate was subtracted and the magnetization was normalized by the volume of the film.

## Conclusion

In summary, we demonstrated the combinatorial thin-film synthesis of layered nitrides Fe*M*$N_2$ (*M* = W, Mo) by reactive sputtering and post-deposition ammonolysis. Synchrotron GIWAXS confirms that both material systems can be stabilized in thin-film form, although they exhibit markedly different composition-dependent structural behavior. $FeWN_2$ maintains the layered structure with comparatively high phase purity across a broad composition range, whereas $FeMoN_2$ exhibits a much narrower phase-purity window and preferentially forms at Fe-poor conditions. Azimuthal GIWAXS analysis reveals strong texture evolution in both material systems, with Fe-rich films exhibiting pronounced out-of-plane fiber texture, and compositions closer to stoichiometry evolving toward broader mixed-orientations. Fe K-edge XAS analysis does not show significant change in the average Fe valence states across the investigated composition range but reveals substantial composition-dependent modifications of the local Fe coordination environment, particularly in $FeMoN_2$, indicating stronger sensitivity to local disorder and phase competition. Magnetic measurements additionally reveal weak ferromagnetic-like behavior in Fe-poor $FeWN_2$, suggesting that local disorder within the triangular Fe lattice may partially relieve magnetic frustration. Overall, this work demonstrates that composition strongly governs the phase stability, crystallographic texture, local structure, and electronic/magnetic properties of iron-based ternary layered nitride thin films, and establishes reactive sputtering combined with ammonolysis as an effective pathway for synthesizing N-rich layered nitrides.

## Acknowledgement

This work was authored by the National Laboratory of the Rockies (NLR) for the U.S. Department of Energy (DOE) under Contract No. DE-AC36-08GO28308. Funding was provided by the U.S. Department of Energy, Office of Science, Basic Energy Sciences, Materials Chemistry program. Use of the Stanford Synchrotron Radiation Lightsource, SLAC National Accelerator Laboratory, is supported by the U.S. Department of Energy, Office of Science, Office of Basic Energy Sciences under Contract No. DE-AC02-76SF00515. XAS experiments were carried out at the 10-ID-B beamline of the Advanced Photon Source, a U.S. Department (DOE) Office of Science User Facility operated for the DOE Office of Science by Argonne National Laboratory under Contract No. DE-AC02-06CH11357. Operations at 10-ID-B are further supported by the Materials Research Collaboration Access Team and its member institutions. We would like to thank Dr Joshua Wright for assistance with experiments at 10-ID-B, as well as Dr. Wenhao Sun and Abrar Rauf for insightful discussions about thermodynamic phase stability. The views expressed in the article do not necessarily represent the views of the DOE or the U.S. Government.

## References

[1] A. Kobayashi, T. Maeda, T. Akiyama, T. Kawamura, and Y. Honda, Sputter Epitaxy of Transition Metal Nitrides: Advances in Superconductors, Semiconductors, and Ferroelectrics, Physica Status Solidi (a) **222**, 2400896 (2025).

[2] H. Bell, Y. M. Shy, D. E. Anderson, and L. E. Toth, Superconducting Properties of Reactively Sputtered Thin-Film Ternary Nitrides, Nb–Ti–N and Nb–Zr–N, Journal of Applied Physics **39**, 2797 (1968).

[3] W. R. L. Lambrecht, M. S. Miao, and P. Lukashev, Magnetic properties of transition-metal nitrides, J. Appl. Phys. **97**, 10D306 (2005).

[4] K. Ito, S. Honda, and T. Suemasu, Transition metal nitrides and their mixed crystals for spintronics, Nanotechnology **33**, 062001 (2021).

[5] W. N. Porter, K. K. Turaczy, M. Yu, H. Mou, and J. G. Chen, Transition metal nitride catalysts for selective conversion of oxygen-containing molecules, Chem. Sci. **15**, 6622 (2024).

[6] S. Hund, O. Gómez-Cápiro, K. Dembélé, S. Berendts, T. Lunkenbein, H. Ruland, E. M. Heppke, and M. Lerch, Fe3Mo3N: Crystal Structure, High-Temperature Behavior, and Catalytic Activity for Ammonia Decomposition, Zeitschrift Für Anorganische Und Allgemeine Chemie **649**, e202300152 (2023).

[7] H. Wang, J. Li, K. Li, Y. Lin, J. Chen, L. Gao, V. Nicolosi, X. Xiao, and J.-M. Lee, Transition metal nitrides for electrochemical energy applications, Chem. Soc. Rev. **50**, 1354 (2021).

[8] S. A. Kadam, R. S. Kate, V. M. Peheliwa, S. A. Shingate, C. C. Sta. Maria, and Y.-R. Ma, A comprehensive review on transition metal nitrides electrode materials for supercapacitor:

Syntheses, electronic structure engineering, present perspectives and future aspects, Journal of Energy Storage **72**, 108229 (2023).
[9] S. Atta, U. NarendraKumar, K. V. A. N. P. S. Kumar, D. P. Yadav, and S. Dash, Recent Developments and Applications of TiN-Based Films Synthesized by Magnetron Sputtering, J. of Materi Eng and Perform **32**, 9979 (2023).
[10] B. Biswas and B. Saha, Development of semiconducting ScN, Phys. Rev. Mater. **3**, 020301 (2019).
[11] *Synthesis and Characterization of Superconducting Thin Films*, in *Thin Films*, Vol. 28 (Elsevier, 2001), pp. 227–270.
[12] M. Stavrev, D. Fischer, C. Wenzel, K. Drescher, and N. Mattern, Crystallographic and morphological characterization of reactively sputtered Ta, TaN and TaNO thin films, Thin Solid Films **307**, 79 (1997).
[13] S. H. Elder, L. H. Doerrer, F. J. DiSalvo, J. B. Parise, D. Guyomard, and J. M. Tarascon, Lithium molybdenum nitride (LiMoN2): the first metallic layered nitride, Chem. Mater. **4**, 928 (1992).
[14] A. Kuhn, M. Martín-Gil, S. Kaskel, J. Strähle, and F. García-Alvarado, The effect of cationic disordering on the electrochemical performances of the layered nitrides LiWN2 and Li0.84W1.16N2, Journal of the European Ceramic Society **27**, 4199 (2007).
[15] X. Ming and X. Kuang, Phase engineering of nitride thin films, Nat. Synth **3**, 1444 (2024).
[16] A. Zakutayev, M. Jankousky, L. Wolf, Y. Feng, C. L. Rom, S. R. Bauers, O. Borkiewicz, D. A. LaVan, R. W. Smaha, and V. Stevanovic, Synthesis pathways to thin films of stable layered nitrides, Nat. Synth **3**, 1471 (2024).
[17] S. Wang, Z. Wang, Z. Zuo, D. Xu, Y. Yan, Z. Feng, and Z. Zeng, Rockseline lead chalcogenides: An alternative class of high-performance thermoelectric materials, Appl. Phys. Lett. **127**, 102110 (2025).
[18] C. L. Rom, R. W. Smaha, C. A. Knebel, K. N. Heinselman, J. R. Neilson, S. R. Bauers, and A. Zakutayev, Bulk and film synthesis pathways to ternary magnesium tungsten nitrides, Journal of Materials Chemistry C **11**, 11451 (2023).
[19] B. Julien, I. A. Leahy, R. W. Smaha, J. S. Mangum, C. L. Perkins, S. R. Bauers, and A. Zakutayev, Structural stability, elemental ordering, and transport properties of layered ${\mathrm{ScTaN}}_{2}$, Phys. Rev. Mater. **9**, 093403 (2025).
[20] K. Katsumata, H. A. Katori, S. Kimura, Y. Narumi, M. Hagiwara, and K. Kindo, Phase transition of a triangular lattice Ising antiferromagnet ${\text{FeI}}_{2}$, Phys. Rev. B **82**, 104402 (2010).
[21] R. Zhong, S. Guo, and R. J. Cava, Frustrated magnetism in the layered triangular lattice materials ${\mathrm{K}}_{2}\mathrm{Co}{({\mathrm{SeO}}_{3})}_{2}$ and ${\mathrm{Rb}}_{2}\mathrm{Co}{({\mathrm{SeO}}_{3})}_{2}$, Phys. Rev. Mater. **4**, 084406 (2020).
[22] D. S. Bem, C. M. Lampe-Önnerud, H. P. Olsen, and H.-C. zur Loye, Synthesis and Structure of Two New Ternary Nitrides: FeWN2 and MnMoN2, Inorg. Chem. **35**, 581 (1996).
[23] D. S. Bem, H. P. Olsen, and H.-C. zur Loye, Synthesis, Electronic and Magnetic Characterization of the Ternary Nitride (Fe0.8Mo0.2)MoN2, Chem. Mater. **7**, 1824 (1995).
[24] S. R. Bauers et al., Ternary nitride semiconductors in the rocksalt crystal structure, Proceedings of the National Academy of Sciences **116**, 14829 (2019).
[25] A. Zakutayev, S. R. Bauers, and S. Lany, Experimental Synthesis of Theoretically Predicted Multivalent Ternary Nitride Materials, Chem. Mater. **34**, 1418 (2022).
[26] A. Miura, X.-D. Wen, H. Abe, G. Yau, and F. J. DiSalvo, Non-stoichiometric Fe*x*WN2: Leaching of Fe from layer-structured FeWN2, Journal of Solid State Chemistry **183**, 327 (2010).
[27] J. A. Steele et al., How to GIWAXS: Grazing Incidence Wide Angle X-Ray Scattering Applied to Metal Halide Perovskite Thin Films, Advanced Energy Materials **13**, 2300760 (2023).

[28] J. C. Fischer, C. Li, S. Hamer, L. Heinke, R. Herges, B. S. Richards, and I. A. Howard, GIWAXS Characterization of Metal–Organic Framework Thin Films and Heterostructures: Quantifying Structure and Orientation, Advanced Materials Interfaces **10**, 2202259 (2023).
[29] G. A. Waychunas, Grazing-incidence X-ray Absorption and Emission Spectroscopy, Reviews in Mineralogy and Geochemistry **49**, 267 (2002).
[30] G. Dräger, R. Frahm, G. Materlik, and O. Brümmer, On the Multipole Character of the X-Ray Transitions in the Pre-Edge Structure of Fe K Absorption Spectra. An Experimental Study, Physica Status Solidi (b) **146**, 287 (1988).
[31] J. Ilavsky, Nika: software for two-dimensional data reduction, J Appl Cryst **45**, 324 (2012).
[32] B. H. Toby and R. B. Von Dreele, GSAS-II: the genesis of a modern open-source all purpose crystallography software package, Journal of Applied Crystallography **46**, 544 (2013).
[33] B. Ravel and M. Newville, ATHENA and ARTEMIS: interactive graphical data analysis using IFEFFIT, Phys. Scr. **2005**, 1007 (2005).
[34] H. Funke, A. C. Scheinost, and M. Chukalina, Wavelet analysis of extended x-ray absorption fine structure data, Phys. Rev. B **71**, 094110 (2005).
[35] H. Ofuchi, T. Matsumoto, and T. Honma, Construction of XAFS standard sample database at BL14B2 in SPring-8, Radiation Physics and Chemistry **218**, 111581 (2024).
[36] B. Julien, A. Rauf, L. A. V. Nagle-Cocco, R. W. Smaha, W. Sun, A. Zakutayev, and S. R. Bauers, Thin-Film Stabilization and Magnetism of $\eta$-Carbide Type Iron Nitrides, (n.d.).
[37] B. Cao, G. M. Veith, J. C. Neuefeind, R. R. Adzic, and P. G. Khalifah, Mixed Close-Packed Cobalt Molybdenum Nitrides as Non-noble Metal Electrocatalysts for the Hydrogen Evolution Reaction, J. Am. Chem. Soc. **135**, 19186 (2013).
[38] D. S. Bem, H. P. Olsen, and H.-C. zur Loye, Synthesis, Electronic and Magnetic Characterization of the Ternary Nitride (Fe0.8Mo0.2)MoN2, Chem. Mater. **7**, 1824 (1995).
[39] K. Hoang and M. Johannes, Defect Physics and Chemistry in Layered Mixed Transition Metal Oxide Cathode Materials: (Ni,Co,Mn) vs (Ni,Co,Al), Chem. Mater. **28**, 1325 (2016).
[40] P. Jivanantaka, N. Monarumit, and S. Satitkune, Oxidation state of Fe in irradiated sapphire samples by XAS technique, J. Phys.: Conf. Ser. **1144**, 012016 (2018).
[41] J. M. Chen, S. C. Haw, J. M. Lee, S. A. Chen, K. T. Lu, M. J. Deng, S. W. Chen, H. Ishii, N. Hiraoka, and K. D. Tsuei, Electronic structure and characteristics of Fe 3d valence states of Fe1.01Se superconductors under pressure probed by x-ray absorption spectroscopy and resonant x-ray emission spectroscopy, J. Chem. Phys. **137**, 244702 (2012).
[42] J. D. Houmes, D. S. Bem, and H.-C. Z. Loye, Synthesis of New Nitrides Using Solid State Oxide Precursors, MRS Online Proceedings Library (OPL) **327**, 153 (1993).
[43] A. Miura, T. Takei, N. Kumada, E. Magome, C. Moriyoshi, and Y. Kuroiwa, Crystal structures and ferromagnetism of FexWN2 (x~0.74, 0.90) with defective iron triangular lattice, Journal of Alloys and Compounds **593**, 154 (2014).

# Composition-dependent Thin-film Synthesis of Layered Ternary Iron Nitrides Fe*M*$N_2$ (*M* = W, Mo)

Baptiste Julien[1]*, Yuwei Yang[2], Liam A. V. Nagle-Cocco[3], Nicholas A. Strange[3], Nicholas M. Bedford[4,5], Andriy Zakutayev[1]*

## Supplementary Materials

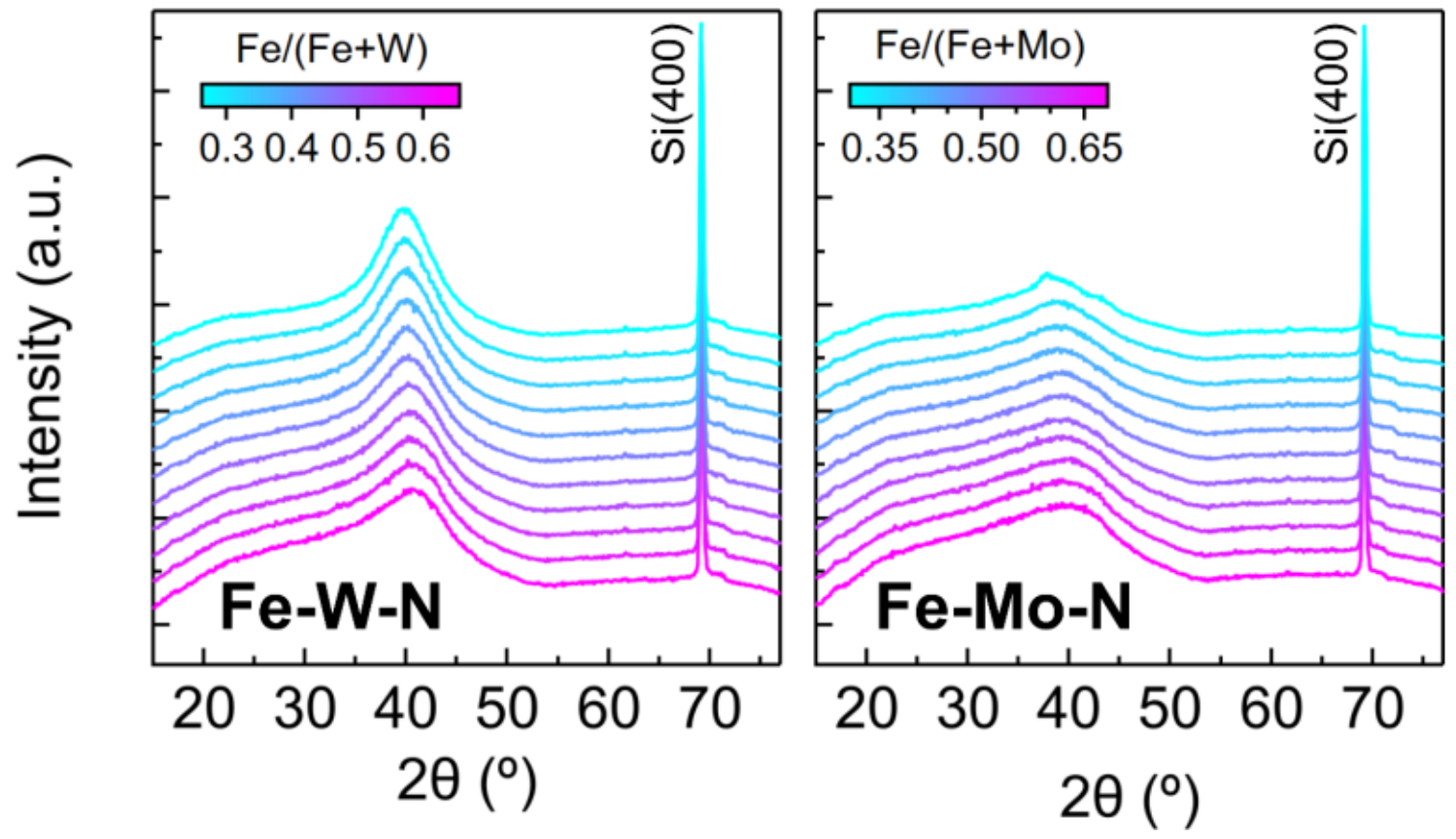

**Figure S1**. Laboratory X-ray diffraction (θ-2θ geometry, λ = 1.54 Å) of pre-annealed Fe-W-N and Fe-Mo-N combinatorial films deposited on Si/$SiN_x$ substrate.

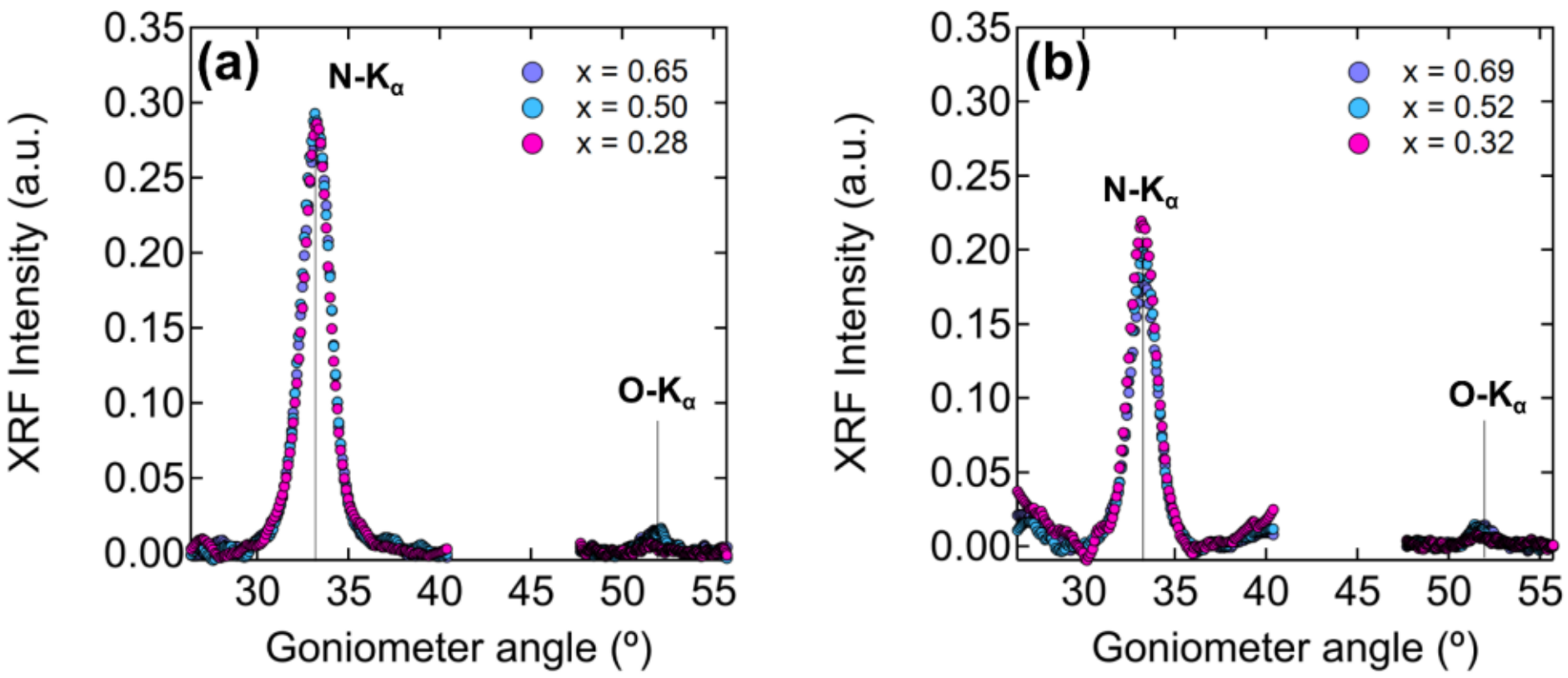

**Figure S2**. WD-XRF spectra of $NH_3$ annealed (a) Fe-W-N and (b) Fe-Mo-N films showing N-$K_\alpha$ and O-$K_\alpha$ emission lines, taken at different cation composition $x$ = Fe/(Fe+*M*) where *M* = W, Mo.

**Table S1**. Phase fraction calculation across composition for Fe-W-N films. For each identified phase, a characteristic reflection is used, and its intensity is normalized to its calculated intensity and compared to the other phases. The intensities were extracted from fully azimuthally-integrated GIWAXS patterns at each composition. The experimental peak intensity $I^{exp}$ corresponds to the peak area, fitted using a Voigt function. The calculated peak intensities $I^{calc}$ were calculated in CrystalDiffract with λ = 0.976 Å and taken relative to the most intense peak (i.e. 100). The phase fraction is calculated from Eq. (1) in the main text.

| **Fe fraction $x$** | **Phase** | **XRD peak** | $I^{exp}$ | $I^{calc}$ | **Phase fraction (%)** |
|---|---|---|---|---|---|
| 0.28 | $FeWN_2$ | (104) | 0.472 | 100 | 75 |
| | $Fe_3N$ | (111) | 0 | 100 | 0 |
| | $W_3N_4$ | (111) | 0.157 | 100 | 25 |
| 0.35 | $FeWN_2$ | (104) | 0.553 | 100 | 93 |
| | $Fe_3N$ | (111) | 0 | 100 | 0 |
| | $W_3N_4$ | (111) | 0.041 | 100 | 7 |
| 0.43 | $FeWN_2$ | (104) | 0.542 | 100 | 100 |
| | $Fe_3N$ | (111) | 0 | 100 | 0 |
| | $W_3N_4$ | (111) | 0 | 100 | 0 |
| 0.50 | $FeWN_2$ | (104) | 0.498 | 100 | 100 |
| | $Fe_3N$ | (111) | 0 | 100 | 0 |
| | $W_3N_4$ | (111) | 0 | 100 | 0 |
| 0.58 | $FeWN_2$ | (104) | 0.5 | 100 | 100 |
| | $Fe_3N$ | (111) | 0 | 100 | 0 |
| | $W_3N_4$ | (111) | 0 | 100 | 0 |
| 0.65 | $FeWN_2$ | (104) | 0.463 | 100 | 93 |
| | $Fe_3N$ | (111) | 0.034 | 100 | 7 |
| | $W_3N_4$ | (111) | 0 | 100 | 0 |

**Table S2**. Phase fraction calculation across composition for Fe-Mo-N films.

| **Fe fraction $x$** | **Phase** | **XRD reflection** | $I^{exp}$ | $I^{calc}$ | **Phase fraction (%)** |
|---|---|---|---|---|---|
| 0.32 | $FeMoN_2$ | (104) | 0.476 | 100 | 75 |
| | Fe | (110) | 0 | 100 | 0 |
| | $Fe_4N$ | (111) | 0 | 100 | 0 |
| | MoN | (002) | 0.07 | 44 | 25 |
| 0.36 | $FeMoN_2$ | (104) | 0.465 | 100 | 81 |
| | Fe | (110) | 0 | 100 | 0 |
| | $Fe_4N$ | (111) | 0 | 100 | 0 |
| | MoN | (002) | 0.048 | 44 | 19 |
| 0.40 | $FeMoN_2$ | (104) | 0.395 | 100 | 100 |
| | Fe | (110) | 0 | 100 | 0 |
| | $Fe_4N$ | (111) | 0 | 100 | 0 |
| 0.48 | $FeMoN_2$ | (104) | 0.377 | 100 | 90 |
| | Fe | (110) | 0.028 | 100 | 7 |
| | $Fe_4N$ | (111) | 0.012 | 100 | 3 |
| 0.56 | $FeMoN_2$ | (104) | 0.256 | 100 | 74 |
| | Fe | (110) | 0.05 | 100 | 15 |
| | $Fe_4N$ | (111) | 0.038 | 100 | 11 |
| 0.63 | $FeMoN_2$ | (104) | 0.22 | 100 | 60 |
| | Fe | (110) | 0.093 | 100 | 26 |
| | $Fe_4N$ | (111) | 0.051 | 100 | 14 |
| 0.69 | $FeMoN_2$ | (104) | 0.215 | 100 | 53 |
| | Fe | (110) | 0.117 | 100 | 29 |
| | $Fe_4N$ | (111) | 0.071 | 100 | 18 |

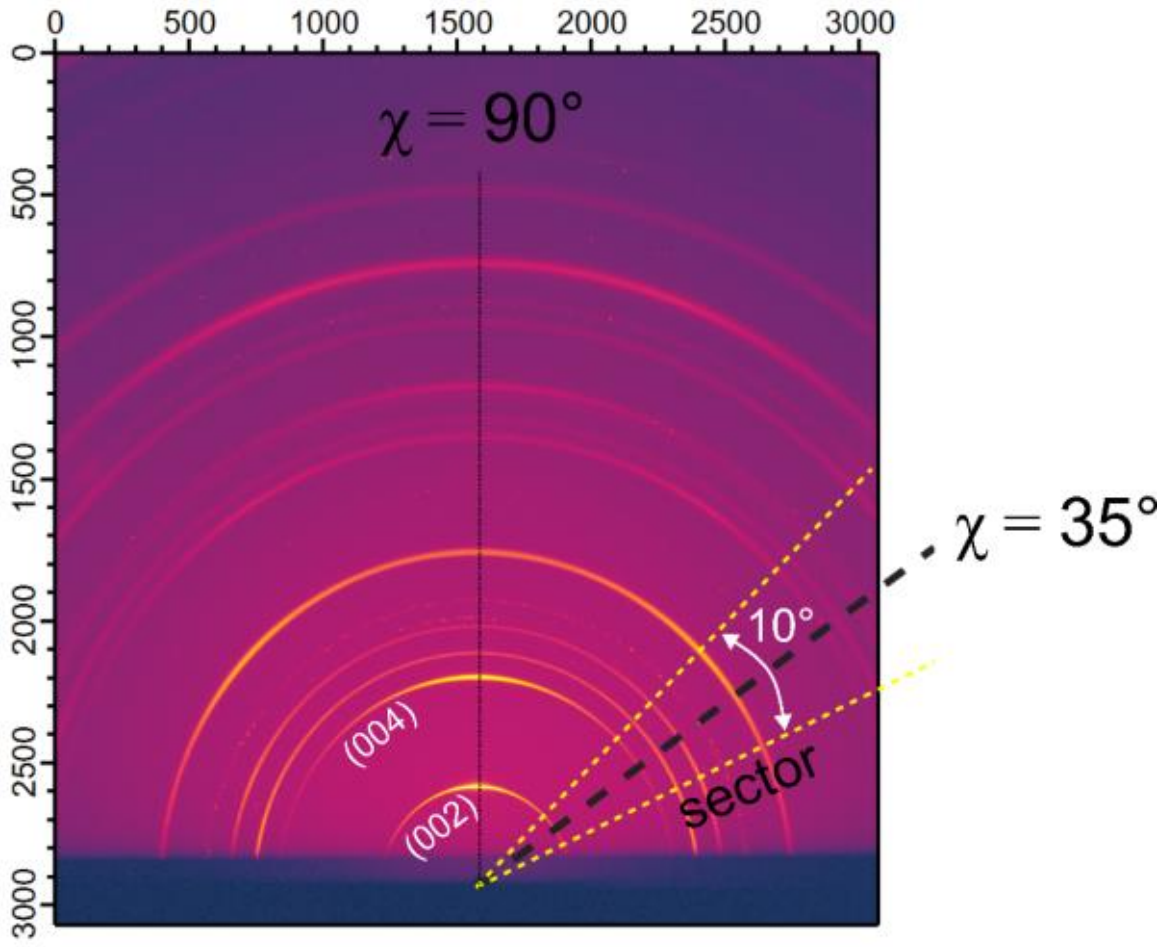

**Figure S3**. Example of a raw GIWAXS image of a Fe-rich $FeWN_2$ film with Fe fraction $x$ = 0.65, showing out-of-plane texture. The image was captured on a Rayonix MX225 detector at SSRL beamline 11-3. Because the area detector is a flat surface, each pixel still captures intensity but due to the Ewald sphere curvature, a region in reciprocal space near $q_z$ axis (i.e. near $q_{xy}$ = 0) is never fully probed in grazing-incidence geometry, especially at high angles, which results in a "missing wedge" in converted image in reciprocal space. For films with preferential orientation along the out-of-plane direction ($q_z$), this can lead to misleading quantification when comparing integrated intensities of crystallographic planes oriented in the $q_z$ direction. In our case, (002) is only weakly affected due to its low angle whereas (004) suffers more significantly from this effect. Thus, it can be necessary to restrict the sector of integration to avoid the $q_z$ direction. The region indicated by the yellow dash lines highlights the azimuthal sector (centered at 35° with a width of 10°) used for intensity integration in order to extract (002) and (004) intensities while minimizing the reciprocal-space inaccessibility effect along $q_z$.

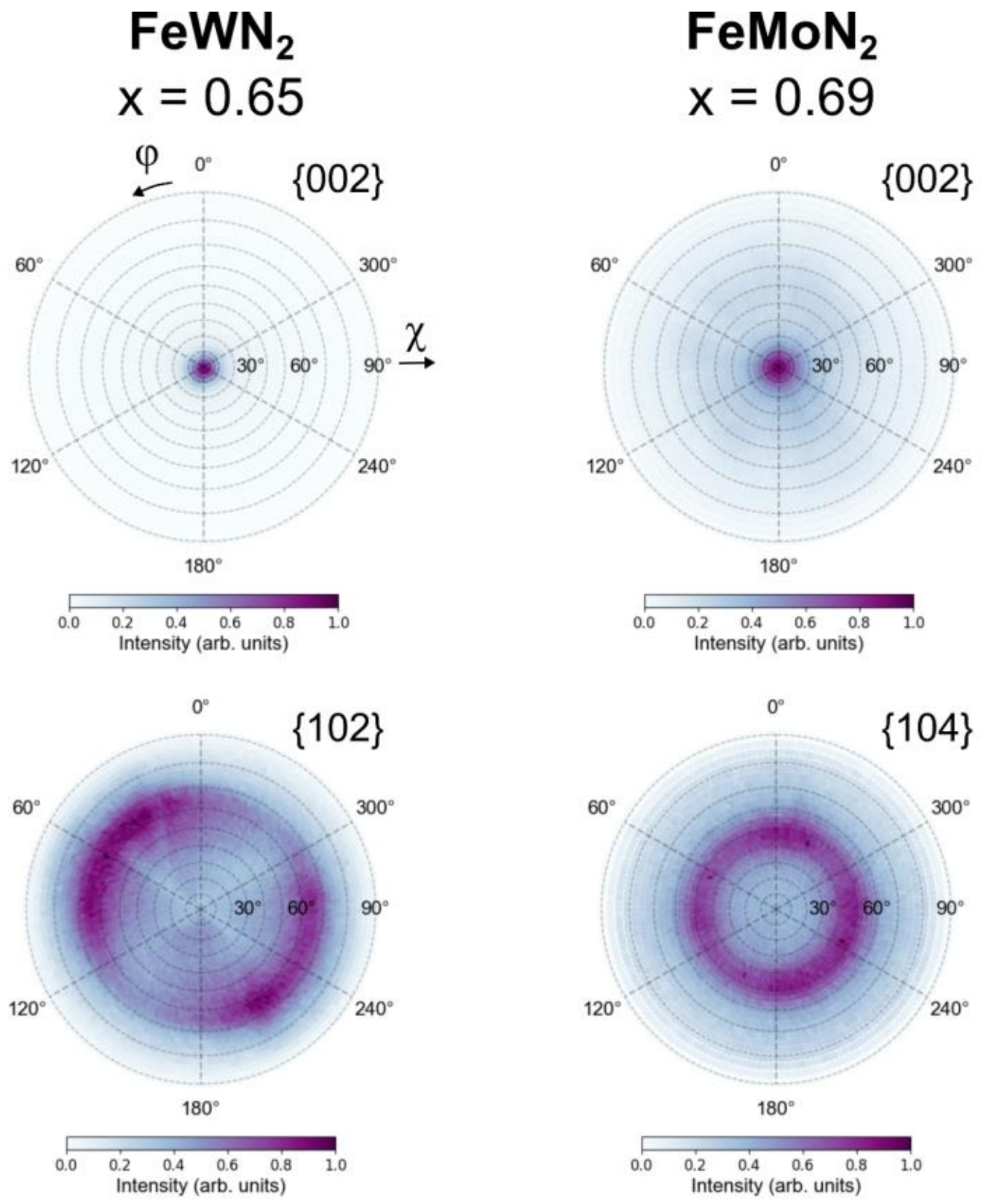

**Figure S4**. XRD pole figures of $FeWN_2$ and $FeMoN_2$ films with $x$ = 0.65 and $x$ = 0.69, respectively, showing (002) reflection and (102)/(104) reflection. We note that in the convention used here, $\chi$ = 0° is equivalent to $\chi$ = 90° in the GIWAXS azimuthal analysis. The sharp central spot in out-of-plane (002) intensity and the intensity ring observed in the in-plane (102) or (104) reflection confirm the fiber-like texture where the grains are preferentially oriented along the c-axis (fiber axis) but are randomly oriented around this axis.

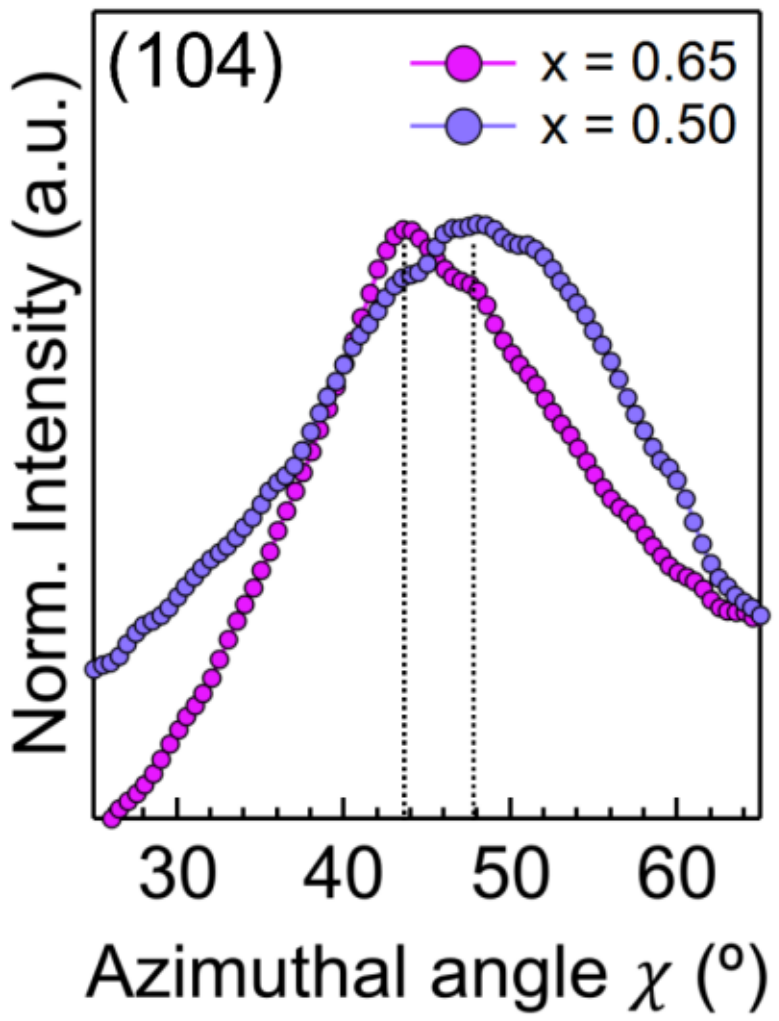

**Figure S5**. Azimuthal intensity profile of (104) reflection in $FeWN_2$ showed at two compositions $x$ = 0.65 and $x$ = 0.50 and zoomed around one maximum. At $x$ = 0.65, the film is preferentially oriented along [001] direction, the peak maximum in $\chi$ correspond to the interplanar between (002) and (104), which is ~48° from calculated structure (corresponding to $\chi$ = 90 – 47 = 43° in our convention). However, at $x$ = 0.50 the preferential orientation shifts toward a more dominant in-plane orientation along [100] direction, and the peak maximum in $\chi$ slightly shifts because the interplanar angle between (001) and (104) is ~42°, corresponding to $\chi$ = 48° in our convention. The presence of a shoulder might indicate a mixed-orientation state where a small fraction of crystallite remains in the other orientation.

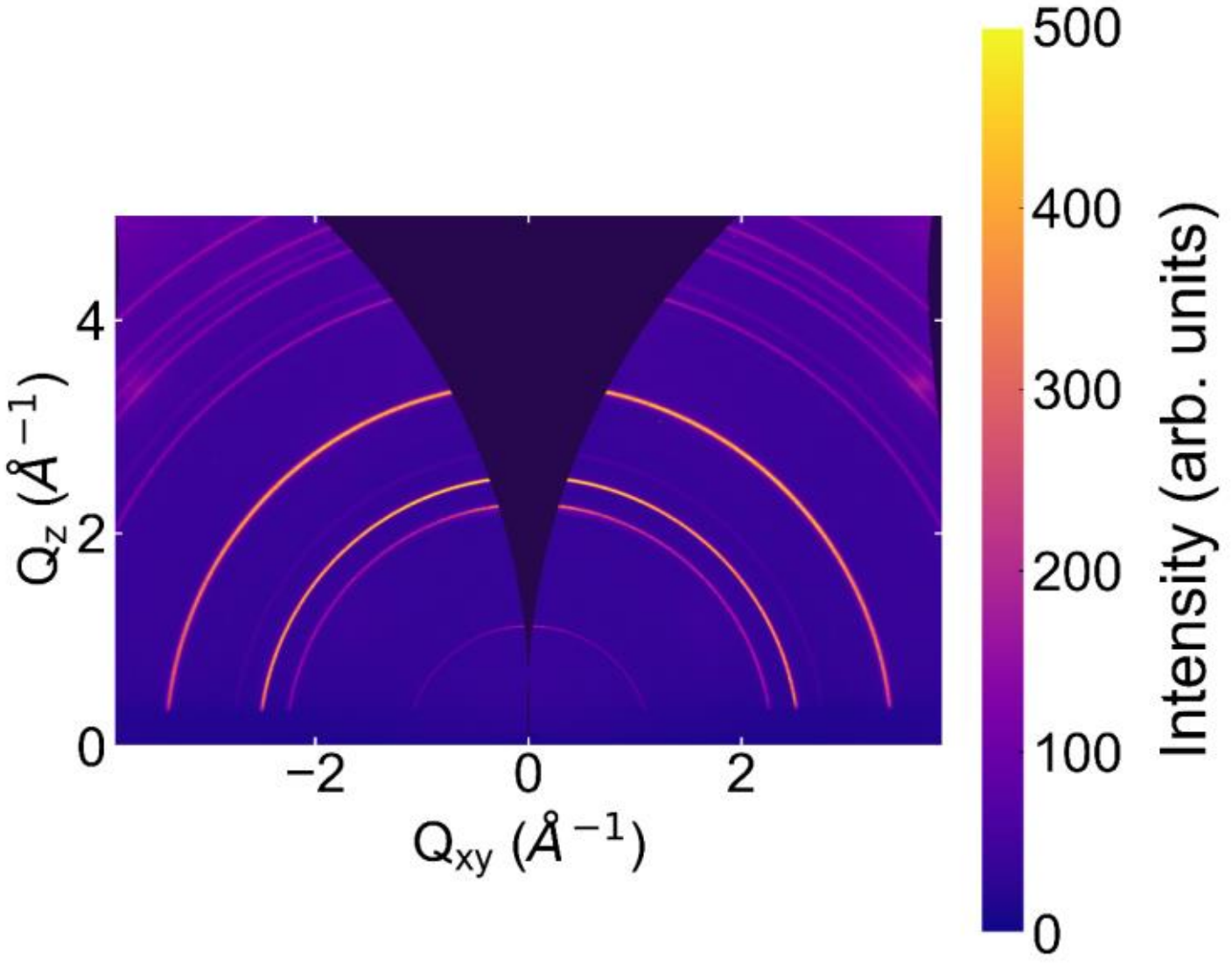

**Figure S6**. GIWAXS reciprocal-space images (Ewald sphere corrected) of phase-pure $FeMoN_2$ at x = Fe/(Fe+Mo) = 0.4, corresponding to $Fe_{0.8}Mo_{1.2}N_2$ stoichiometry. The isotropic diffraction rings indicates good overall polycrystallinity without strong texture effect.

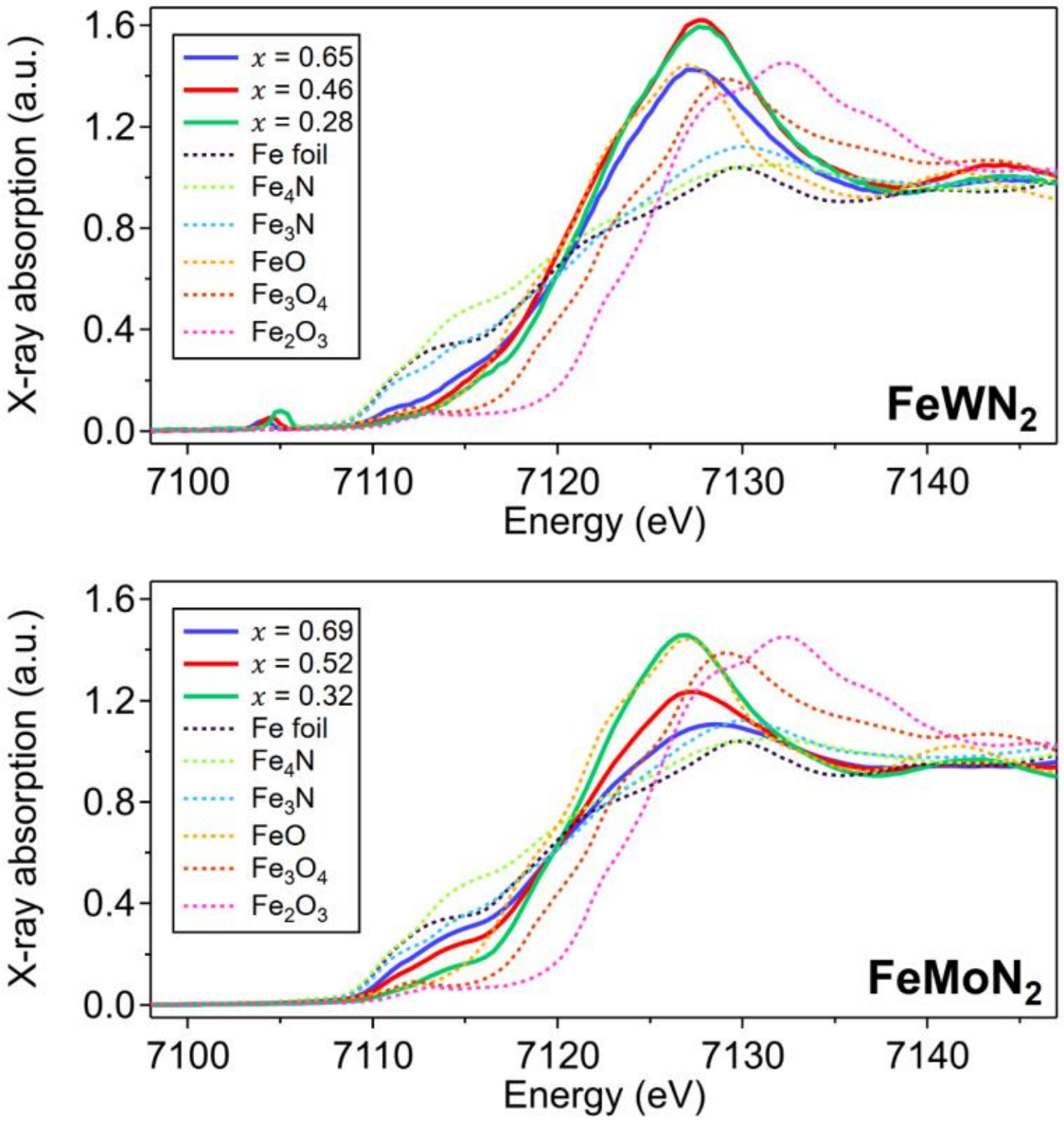

**Figure S7**. Fe K-edge XANES spectra for $FeWN_2$ and $FeMoN_2$ at different Fe fraction $x$. Various Fe-based standard XANES spectra were calibrated and added for oxidation state comparison (from BL14B2 XAFS Standard Sample Database [1]). For all composition, the Fe-K edge of Fe(W,Mo)$N_2$ is shifted to higher energy with respect to Fe, $Fe_4N$ and $Fe_4N$, indicating a more electron-deficient Fe environment. However, the edges remains below those of $Fe_3O_4$ (where Fe presumably has an effective oxidation state of +2.6) and $Fe_2O_3$ (Fe +3), and coincide with FeO, consistent with a low positive $Fe^{2+}$-like oxidation state.

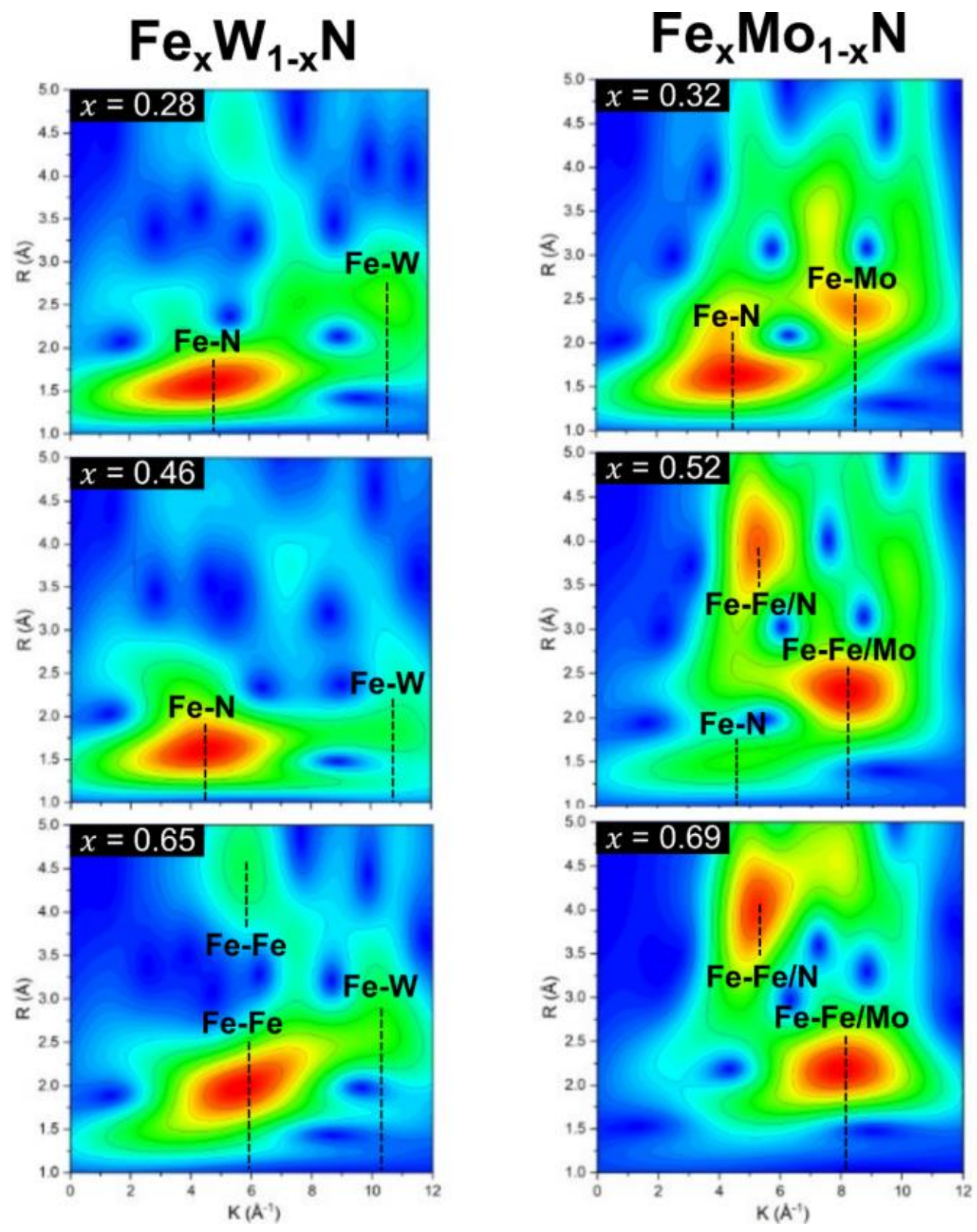

**Figure S8**. Wavelet-transformed EXAFS contour plots of $FeWN_2$ and $FeMoN_2$ at different Fe fraction $x$.

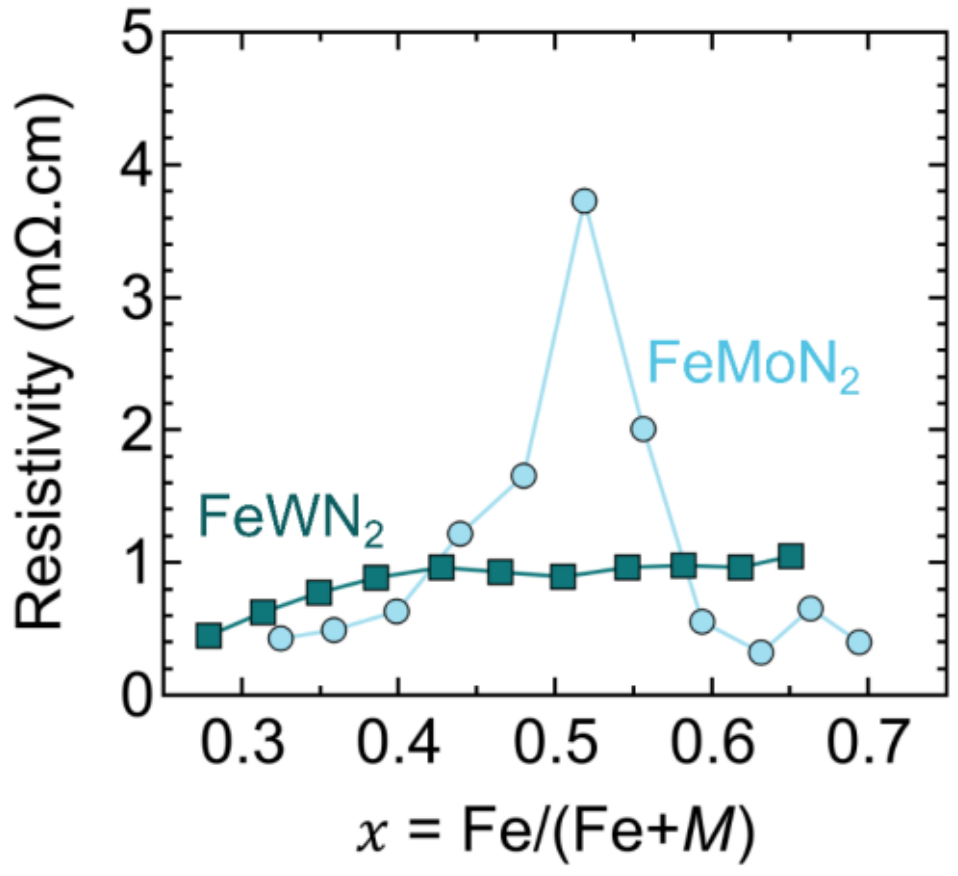

**Figure S9**. Composition-dependent electrical resistivity of post-annealed $FeWN_2$ and $FeMoN_2$ films. The sheet resistivity was measured by 4 points probe method and the resistivity was extracted from the local thickness estimated by XRF.

[1] H. Ofuchi, T. Matsumoto, and T. Honma, Construction of XAFS standard sample database at BL14B2 in SPring-8, Radiation Physics and Chemistry **218**, 111581 (2024).